\documentclass[11pt,twocolumn,numbers]{article}

\usepackage{graphicx}
\usepackage{amssymb}
\usepackage{amsthm,amsmath,dsfont}
\usepackage{mathtools}

\usepackage{lineno}

\usepackage[colorlinks=true, citecolor=blue]{hyperref}
\usepackage[linesnumbered,lined,ruled,commentsnumbered]{algorithm2e}
\usepackage{algcompatible}

\usepackage{wrapfig}
\usepackage[numbers,sort&compress]{natbib}
\usepackage{comment}
\usepackage{longtable}
\usepackage{multirow}
  
\usepackage[bottom]{footmisc}
\usepackage{booktabs}

\usepackage{float}
\usepackage{subfig}
\usepackage{wrapfig}
\usepackage{framed,xcolor}
\usepackage{newfloat}
\usepackage[xcolor,framemethod=TikZ]{mdframed}
\usepackage{mathpazo}
\usepackage{amsthm}
\usepackage{thmtools}
\usepackage{authblk}
\patchcmd{\thebibliography}{\section*{\refname}}{}{}{}
\usepackage{widetext}
\usepackage{tikz}
\usetikzlibrary{arrows,shapes,chains,calc,patterns,decorations.pathreplacing}
\usepackage{stmaryrd}
\usepackage{fancyhdr}

\renewcommand{\headrulewidth}{2pt}

\newlength\FHoffset

\fancyheadoffset{\FHoffset}

\newlength\FHleft
\newlength\FHright

\newbox\FHline
\setbox\FHline=\hbox{\hsize=\paperwidth%
	\hspace*{\FHleft}%
	\rule{\dimexpr\headwidth-\FHleft-\FHright\relax}{\headrulewidth}\hspace*{\FHright}%
}

\newcommand{\RR}{\mathbb{R}}
\newcommand{\ZZ}{\mathbb{Z}}
\newcommand{\GX}{\mathcal{G}}

\newcommand{\VX}{\mathcal{V}}
\newcommand{\LX}{\mathcal{L}}
\newcommand{\AX}{\mathcal{A}}
\newcommand{\AXI}{\mathcal{A}_{\VX/S}}
\newcommand{\AXJ}{\mathcal{A}_{\mathsf{conf}}}
\newcommand{\TX}{\mathcal{T}}
\newcommand{\CX}{\mathcal{C}}
\newcommand{\CO}{\mathcal{C}_{\mathsf{O}}}
\newcommand{\CR}{\mathcal{C}_{\mathsf{R}}}
\newcommand{\CS}{\mathcal{C}_{\mathsf{S}}}
\newcommand{\BX}{\mathcal{B}}
\newcommand{\BD}{\mathcal{B}_{\mathsf{D}}}
\newcommand{\BM}{\mathcal{B}_{\mathsf{M}}}

\newtheoremstyle{theoremdd}
{\topsep}
{\topsep}
{\itshape}
{0pt}
{\fontfamily{cmss}\selectfont\bfseries}
{.}
{ }
{\thmname{#1}\thmnumber{ #2}\thmnote{ (#3)}}

\theoremstyle{theoremdd}

\mdfdefinestyle{myStyle}{roundcorner=5pt,backgroundcolor=brown!20,linecolor=red!40!black,linewidth=1.5pt}

\usepackage{titlesec}
\titleformat*{\section}{\fontfamily{cmss}\selectfont\large\bfseries\color{red!40!black}}
\titleformat*{\subsection}{\fontfamily{cmss}\selectfont\normalsize\bfseries\color{red!40!black}}
\titleformat*{\subsubsection}{\fontfamily{cmss}\selectfont\normalsize\color{red!40!black}}
\usepackage[left=0.75 in, right=0.75 in, top=1.1 in, bottom=1.3 in]{geometry}
\graphicspath{{./Figures/}}

\newcommand\blfootnote[1]{%
	\begingroup
	\renewcommand\thefootnote{}\footnote{#1}%
	\addtocounter{footnote}{-1}%
	\endgroup
}
\renewcommand\abstractname{\fontfamily{cmss}\selectfont\normalsize\bfseries\color{red!40!black}\textbf{Abstract}}

\renewenvironment{abstract}{%
	\centering\small
	\list{}{\leftmargin1.5cm \rightmargin\leftmargin}
	\item\relax
	
	\begin{mdframed}[style=myStyle]
		\item[\hskip\labelsep\scshape\abstractname.]%
	}{%
	\end{mdframed}
	\endlist \par\bigskip
}
\begin{document}

		
		
		\title{\fontfamily{cmss}\selectfont\color{red!40!black} Noticeability Versus Impact in Traffic Signal Tampering}
		


\author[1]{\fontfamily{cmss}\selectfont Bilal Thonnam Thodi}
\author[1]{\fontfamily{cmss}\selectfont Timothy Mulumba}
\author[1,2]{\fontfamily{cmss}\selectfont Saif Eddin Jabari$^{\star,}$}

\affil[1]{{ \small New York University Tandon School of Engineering, Brooklyn NY, U.S.A.}}
\affil[2]{{ \small New York University Abu Dhabi, Saadiyat Island, P.O. Box 129188, Abu Dhabi, U.A.E.}}

\date{}


\twocolumn[
\begin{@twocolumnfalse}
	
\maketitle	

\begin{abstract}
	This paper investigates the vulnerability of urban traffic networks to cyber-attacks on traffic lights.  We model traffic signal tampering as a bi-objective optimization problem that simultaneously seeks to reduce vehicular throughput in the network over time (maximize impact) while introducing minimal changes to network signal timings (minimize noticeability).  We represent the Spatio-temporal traffic dynamics as a static network flow problem on a time-expanded graph.  This allows us to reduce the (non-convex) attack problem to a tractable form, which can be solved using traditional techniques used to solve linear network programming problems.  We show that minor but objective adjustments in the signal timings over time can severely impact traffic conditions at the network level. We investigate network vulnerability by examining the concavity of the Pareto-optimal frontier obtained by solving the bi-objective attack problem.  Numerical experiments are carried to illustrate the types of insights that can be extracted from the Pareto-optimal frontier.  For instance, our experiments suggest that the vulnerability of a traffic network to signal tampering is independent of the demand levels.
	
	\medskip
	
	\textbf{\fontfamily{cmss}\selectfont\color{red!40!black} Keywords}: Network vulnerability, signal tampering, traffic flow dynamics, urban traffic networks, adversarial attacks, cyber-security.
\end{abstract}
\bigskip
\end{@twocolumnfalse}
]

		
	
	
	

\section{Introduction}
\label{sec:introduction}
Urban\blfootnote{$^{\star}$ Corresponding author, Email: \url{sej7@nyu.edu}} network control has evolved from standalone control devices into sophisticated cyber-physical systems, where the integration of cyber elements such as sensing devices, communication channels and control units has greatly influenced the physical flow dynamics and the network controllability \cite{reilly2016freeway}. Various state-of-the art adaptive traffic signal control algorithms which rely on these cyber elements have been designed to better manage and control the vehicular flows through these networks \cite{li2019pwbp,lin2019transferable,lin2020pay}. This, however, has introduced vulnerabilities in traffic networks, especially to cyber attacks, an issue that has gained recent attention \cite{feng2018cyberattack,laszka2016sigtamp,ghafouri2016ftvulnera,li2016assessing}.

We study the vulnerability of urban road networks to traffic \emph{signal tampering} attacks \cite{laszka2016sigtamp}, which involves adversarial adjustment of signal settings to negatively impact network-wide traffic operations. For example, an adversary with unauthorized access to the traffic signal control infrastructure can set the traffic signal timings to a sub-optimal configuration so as cause reduced intersection throughput and excessive queuing. This effect, albeit local, has impacts that propagate to neighboring intersections and, over longer periods of time, can cripple an entire network. We specifically consider two critical aspects of such attacks: their \emph{impact} on network traffic conditions and their \emph{noticeability} to the network operators/controllers and other users. Using a bi-criteria optimization framework, we show that such unnoticeable but objective adjustments to the traffic signal settings over a pre-determined time period can create cascading traffic congestion in the urban road network.

Optimizing a dynamical traffic system to achieve the desired (adversarial) objective presents a challenging task. For instance, to model the time-varying network traffic conditions as a function of traffic signal timing changes, it is important to capture the traffic queue build-up and dissipation at each intersection in the network with reasonable level of detail. This renders the mathematical formulation of the optimization problem more complex and, thus, more difficult to solve. Moreover, to incorporate signalized intersection control in the formulation, it is essential to capture interactions between conflicting movements at the intersections. These typically require the use of integer variables, particularly if time is treated as a discrete variable.  Examples of such formulations include \cite{lo2001cbt,lo1999nts,lin2004emi,beard2006sos}, which use large numbers of binary variables, and researchers often resort to meta-heuristic solution methods given such difficulties; see for example \cite{laszka2016sigtamp,grossmann1995mcf}.

The dynamical nature of the problem coupled with the discrete nature of signal control introduces modeling complexities that we overcome with a \emph{super-graph} representation of the problem. Our super-graph is a directed graph, which compactly captures the Spatio-temporal evolution of vehicular flow dynamics of any real-world traffic network. Using reasonable approximations, this equivalent representation of traffic allows us to exploit simple solution techniques that are based on dynamic programming principles.  Moreover, the super-graph representation permit us to convert the bi-objective optimization signal tampering problem to a more tractable form with polynomial complexity. This is based on the total unimodularity property of the adjacency matrices of graphs, for which integer optimal solutions are obtained by solving the trivial linear relaxation of the respective problem \cite{wols1998integerProg,ahuja1993netflows}; further on this in Sec.~\ref{subsec:supergraph}. Finally, we also propose a network vulnerability measure based on the properties of the solution set obtained by solving the bi-objective attack problem. The solution set forms an Pareto-optimal frontier \cite{deb2014multi}, its properties provide critical insights to the vulnerability of the network. In short, we develop a novel static representation of traffic network dynamics as static flows on a static directed graph in Sec.~\ref{sec:networktrafficflow}.  We then present an attack model for urban road network traffic via signal tampering and demonstrate that it can be solved to optimality using classical bi-objective programming techniques that were developed for linear problems in Sec.~\ref{sec:failure-model}.  In Sec.~\ref{sec:numerical-expts}, we conduct numerical experiments on a four toy networks to illustrate some of the insights that can be extracted from the Pareto-optimal frontier that is obtained as an output of the bi-objective attack problem.

\section{Modeling urban network traffic flows}
\label{sec:networktrafficflow}

\subsection{Network traffic dynamics} \label{subsec:trafficdynamics}

\begin{figure*}[ht!]
	\centering
	\resizebox{0.7\textwidth}{!}{%
		\includegraphics{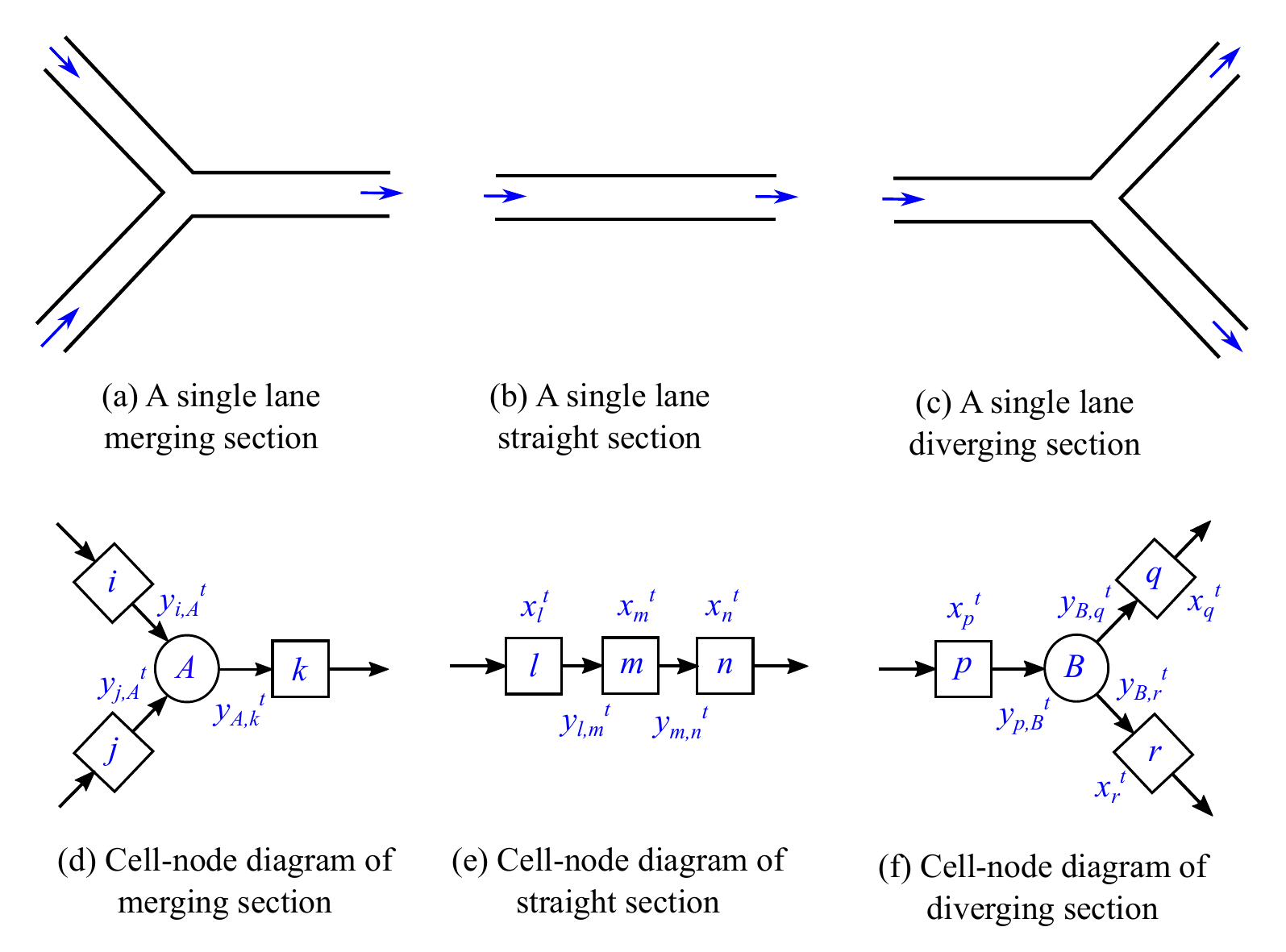}}
	\caption{\small Illustration of the directed graph representation for a simple road network.}
	\label{fig:cell-node-illustr}
\end{figure*}
We use a cell-based traffic flow model to capture the vehicular flow dynamics of an urban road network.  That is, we divide network road segments (links) into smaller sections (cells) to capture the build-up and dissipation of congestion within the network links.  The model assumes space and time to be discrete, and that flows across cell boundaries do not exceed the numbers of vehicles in the sending cells, the capacities of the two cells, or maximum occupancy of the downstream cell. 
Let $\TX \equiv \{0,1,\hdots, |\TX|\}$ denote the set of discrete time steps with a finite horizon $|\TX|$.  We use a uniform discrete time step length of $\Delta \tau$, that is, time $t \in \TX$ represents $t \Delta \tau$ units of time (typically seconds) from the initial time step. Let $\CX$ denote the set of network cells; the length of each cell $\Delta x$ depends on the distance traveled by a vehicle at free flow speed over an interval of length $\Delta \tau$.  For example, if $\Delta \tau = 1$ second and the free-flow speed is 90 km/hr, then the cells are chosen to be $\Delta x = 25$ meters in length and the road segments are discretized accordingly.  We denote by $x_i^t$ the occupancy of cell $i \in \CX$ at time $t \in T$.  We represent the flows between adjacent cells by objects that we refer to as ``connectors'' \cite{zilia2000dso} and denote by $y_{ij}^t$ the number of vehicles using connector $(i,j)$ at time $t \in T$, i.e., the number of vehicles departing cell $i \in \CX$ into \emph{adjacent} cell $j \in \CX$ at time $t$.  We denote the set of connectors in the network by $\LX$.  The flow capacity of cell $i$ is denoted by $Q_i$; this represents the maximal number of vehicles that can flow through the cell in single time step ($\Delta \tau$).  We chose $\Delta \tau$ so that $Q_i=1$ vehicle per $\Delta \tau$ units of time (per lane) and adjust the cell lengths accordingly. We demonstrate below that little generality is lost in making this assumption.  The maximum occupancy of cell $i$ is denoted by $N_i$; this represents the maximal number of vehicles that can be present in the cell in units of vehicles per $\Delta x$ units of distance.

Network intersections are modeled as ordinary cells connected to transshipment nodes.  The difference between nodes and cells is that network nodes only serve as transfer stations and do not hold any vehicles. The set of all such transshipment nodes in the network is denoted by $\BX=\BD \cup \BM$, where $\BD$ and $\BM$ are sets of diverging and merging nodes, respectively; the former are nodes with one inbound connector (one incoming flow) and one or more outbound connectors, the latter are nodes with one or more inbound connectors and only one outbound connector. Intersections with multiple inbound flows and multiple outbound flows are represented by more than one node in $\BD$ and $\BM$ as illustrated below.  For each $i \in \CX \cup \BX$, we denote by $\Gamma(i)^- \subset \CX \cup \BX$ the adjacent network objects that are in the immediate upstream of cell (or transshipment node) $i$.  Similarly, $\Gamma(i)^+ \subset \CX \cup \BX$ is the set of adjacent cells or transshipment nodes that are in the immediate downstream of $i$. This cell-node representation is illustrated in Fig.~\ref{fig:cell-node-illustr} for a typical two intersection system.

\textit{Traffic dynamics.} We divide the set of network cells $\CX$ into disjoint sets of ``ordinary'' cells $\CO$, ``source'' cells $\CR$, and ``sink'' cells $\CS$, i.e., $\CX = \CO \cup \CR \cup \CS$.  Ordinary cells are those on the interior of the network, while source and sink cells are on the boundary of the network. By \emph{boundary}, we mean any object that is adjacent to the exterior of the network, this includes the extremities of the network and interior locations that are connected to interior sources or sinks (such as parking structures).  For $i \in \CO$ the dynamics are given by the mass balance equation
\begin{equation}
x_i^t - x_i^{t-1} - y_{hi}^{t-1} + y_{ij}^{t-1} = 0, \label{eqn_flow_a}
\end{equation}
where $\{h\} = \Gamma(i)^-$ and $\{j\} = \Gamma(i)^+$.  For source cells $i \in \CR$, the dynamics are written as
\begin{equation}
x_i^t - x_i^{t-1} + y_{ij}^{t-1} = d_i^t, \label{eqn_flow_b}
\end{equation}
where $d_i^t$ is the exogenous demand at source cell $i$.  Similarly, for $i \in \CS$, the dynamics follow the mass balance equation
\begin{equation}
x_i^t - x_i^{t-1} - y_{hi}^{t-1} = 0. \label{eqn_flow_c}
\end{equation}
Flows through diverge ($i \in \BD$) and merge ($i \in \BM$) nodes are represented, for each $t \in \TX$, by 
\begin{equation}
\sum_{j \in \Gamma(i)^+} y_{ij}^t - y_{hi}^t = 0 \label{eqn_flow_d}
\end{equation}
and
\begin{equation}
y_{ij}^t - \sum_{h \in \Gamma(i)^-} y_{hi}^t = 0, \label{eqn_flow_e}
\end{equation}
respectively.  To capture the physical constraints on the connector flow, we impose the constraint
\begin{equation} 
y_{ij}^{t} \le \min\{ x_{i}^{t}, Q_i, Q_j, N_j - x_{j}^{t} + y_{jk}^t \}. \label{eqn_flow_res_a} 
\end{equation}
That is, the flow on the connector cannot exceed (i) the number of vehicles in the sending cell $i$
\begin{equation}
y_{ij}^t - x_i^t \le 0, \label{eqn_flow_res_b}
\end{equation}
(ii) the flow capacities of both the sending cell $i$ and the receiving cell $j$ ($Q_{ij} \equiv \min \{Q_i,Q_j\}$):
\begin{equation}
y_{ij}^{t} \le  Q_{ij}, \label{eqn_flow_res_c}
\end{equation}
and (iii) the available space in the receiving cell $j$ (accounting for departing flows $y_{jk}^t$):
\begin{equation}
y_{ij}^{t} + x_{j}^{t} - y_{jk}^t \le N_j. \label{eqn_flow_res_d}
\end{equation}

\begin{figure}[h!]
	\centering
	\subfloat[][Intersection layout]{\resizebox{0.275\textwidth}{!}{
			\includegraphics[width=0.275\textwidth,origin=c]{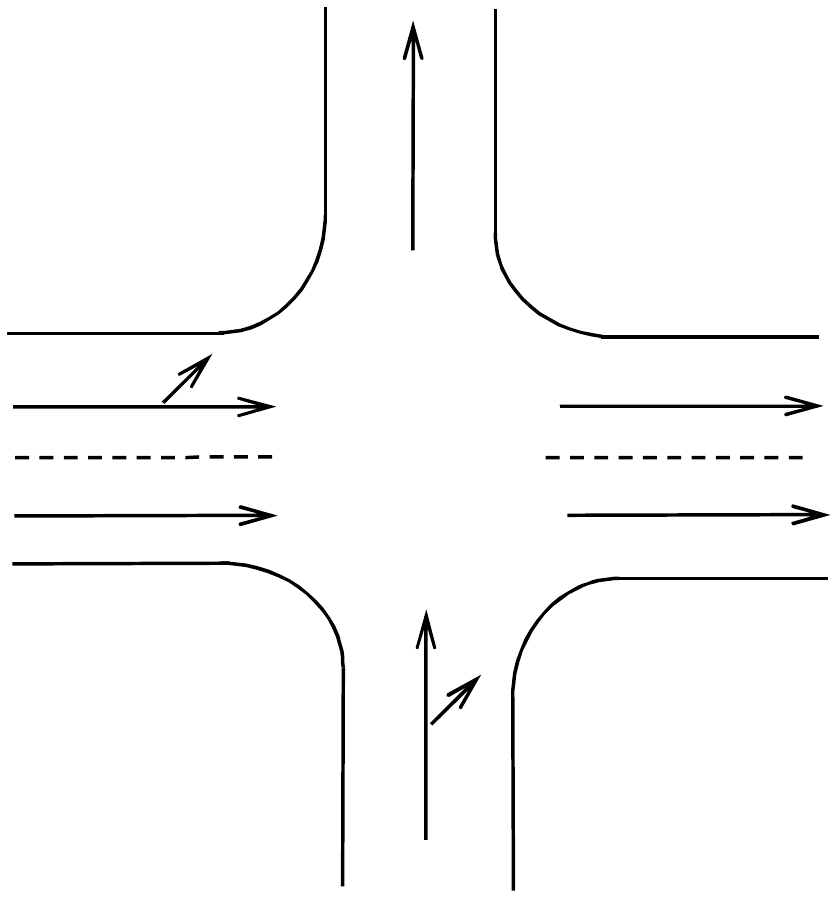}} 
		\label{fig:2lanea}} 	
	
	\subfloat[][Network representation of conflicting movements]{\resizebox{0.35\textwidth}{!}{
			\includegraphics[width=0.35\textwidth,origin=c]{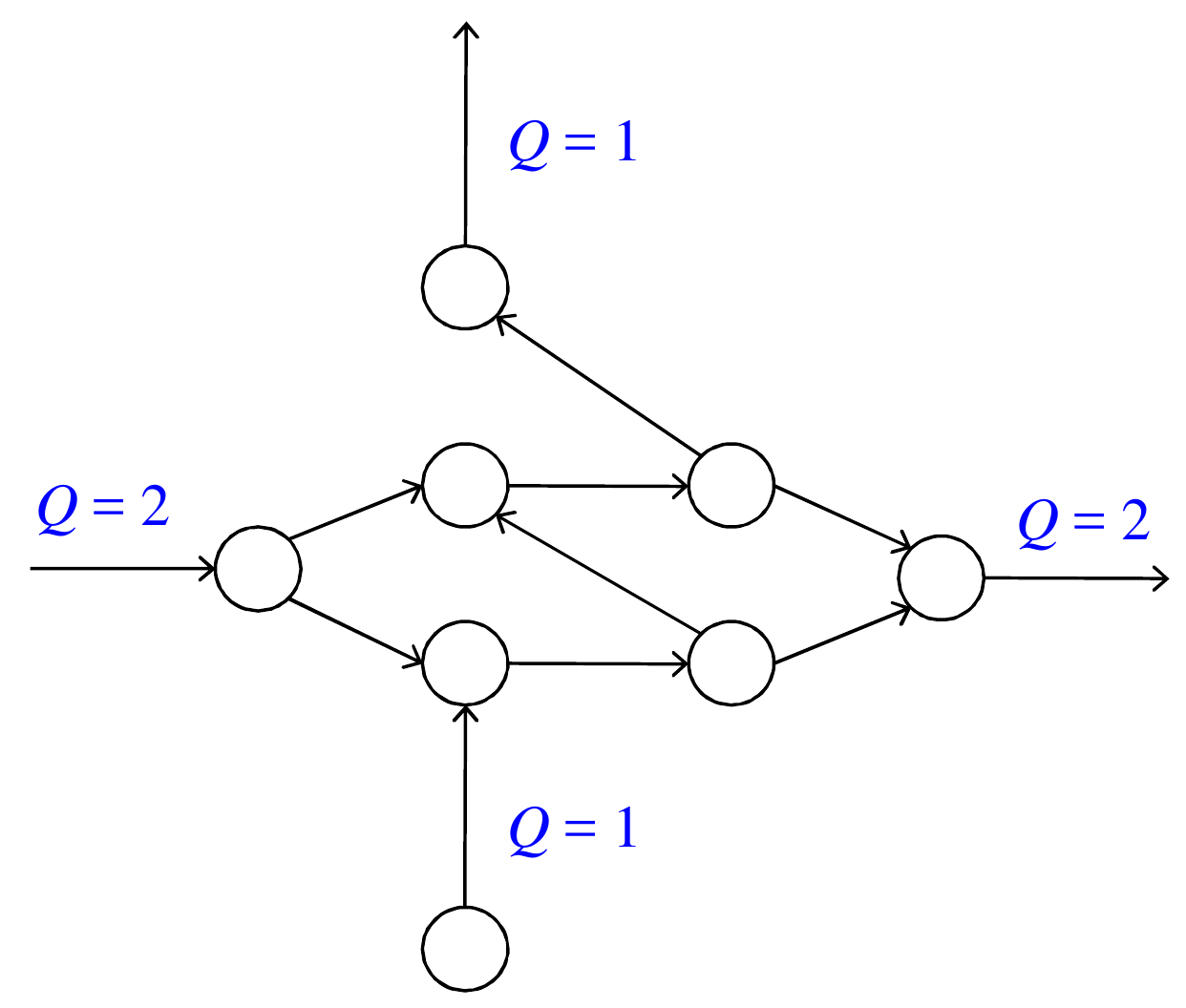}} 
		\label{fig:2laneb}}
	\caption{\small Network representation of conflicting movements for a simple intersection with two lanes running west-east and one lane running south-north.}
	\label{fig:2lane}
\end{figure}

\textit{Intersection conflicting movements.} Additional constraints arise as a result of conflicts at network intersections.  We now show that these constraints can be represented graphically. 
\begin{figure}[h!]
	\centering
	\subfloat[][A pair of conflicting movements]{\resizebox{0.35\textwidth}{!}{
			\includegraphics[width=0.35\textwidth,origin=c]{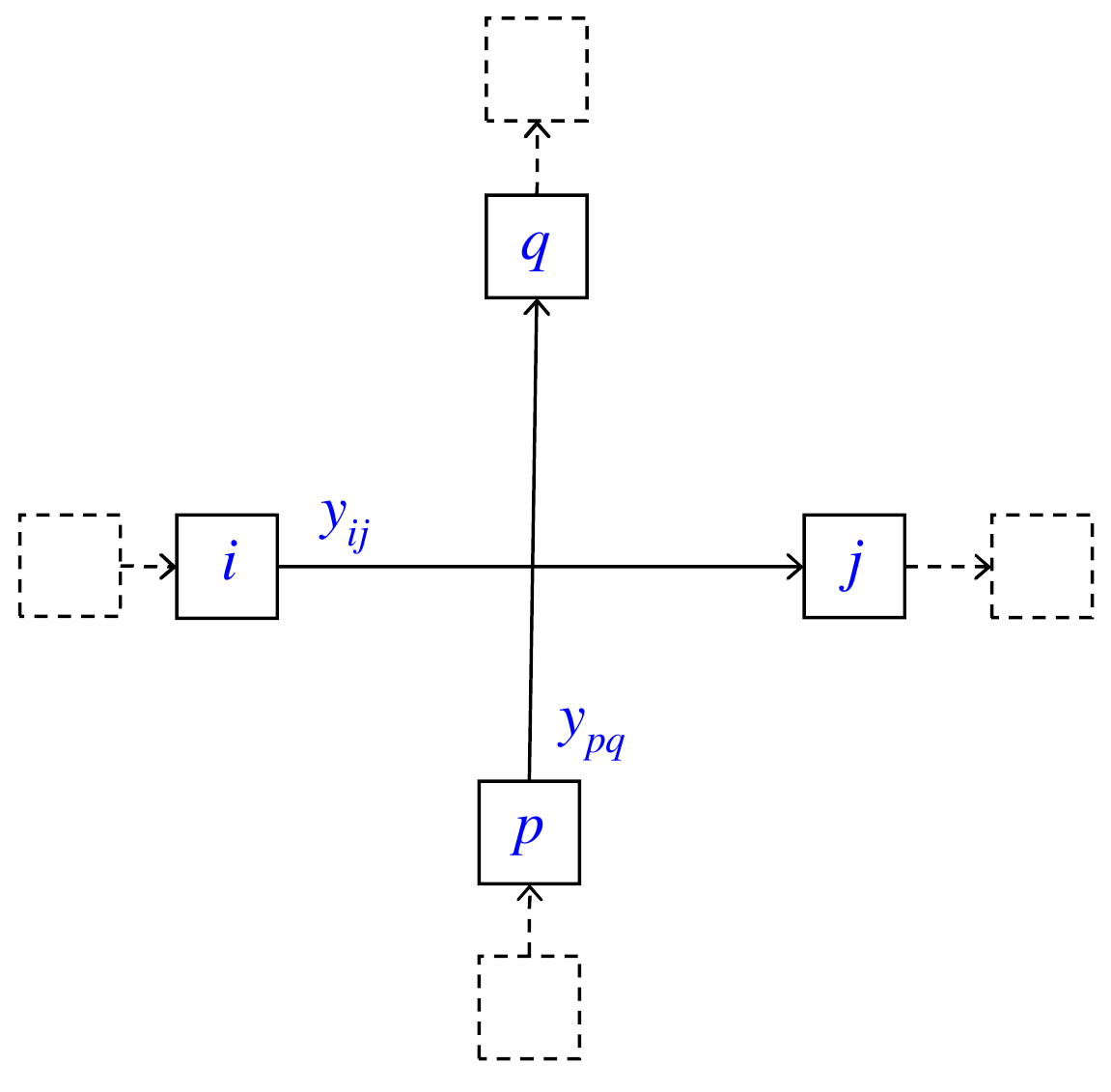}} 
		\label{fig:node-treata}} 	
	
	\subfloat[][Modified representation to capture conflicts]{\resizebox{0.35\textwidth}{!}{
			\includegraphics[width=0.35\textwidth,origin=c]{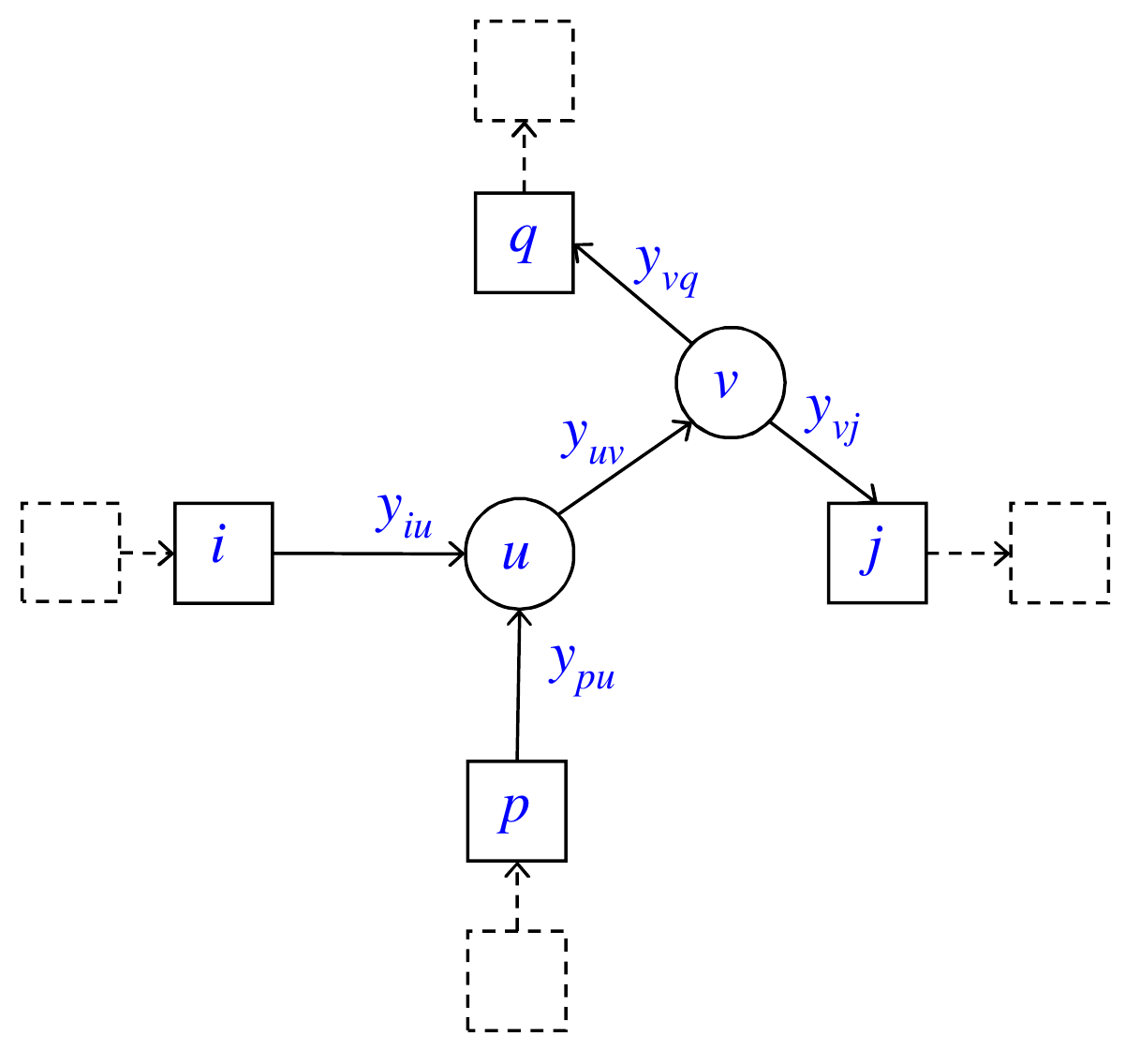}} 
		\label{fig:node-treatb}}
	\caption{\small Modeling conflicting movements.}
	\label{fig:node-treat}
\end{figure}
Consider a pair of intersection movements that conflict with one another, e.g., the two opposing through movements shown in Fig.~\ref{fig:node-treata}.  The two movements are represented by connectors: $(i,j)$ and $(p,q)$.  The conflict dictates that the connector flows cannot be simultaneously positive. To capture this, we can introduce a pair of binary variables $\alpha_{ij}^t,\alpha_{pq}^t \in \{0,1\}$ so that $\alpha_{ij}^t + \alpha_{pq}^t \le 1$ for all $t \in \TX$.  For these two movements, the connector capacity constraints are given by $y_{ij}^{t} \le  Q_{ij} \alpha_{ij}^t$ and $y_{pq}^{t} \le  Q_{pq} \alpha_{pq}^t$.  For the case where both $(i,j)$ and $(p,q)$ have the same (single-lane) flow capacity: $Q_{ij} = Q_{pq} = 1$ vehicle per $\Delta \tau$, these constraints can alternatively be written as
\begin{equation}
y_{ij}^t + y_{pq}^t \le 1 \label{eqn_node_b}
\end{equation}
along with $y_{ij}^t, y_{pq}^t \in \{0,1\}$. In order to represent this graphically, we introduce two transshipment nodes, $u$ and $v$ as depicted in Fig.~\ref{fig:node-treatb} and add the associated mass-balance equations to the system:
\begin{equation}
y_{uv}^{t} - y_{iu}^{t} - y_{pu}^{t} = 0
\end{equation}
and
\begin{equation}
y_{vj}^{t} + y_{vq}^{t} - y_{uv}^{t} = 0
\end{equation}
along with a capacity constraint on the new connector:
\begin{equation}
y_{uv}^{t} \leq 1.
\end{equation}
Connectors $(i,u)$ and $(p,u)$ represent the outbound flows from cells $i$ and $p$, while connectors $(v,j)$ and $(v,q)$ represent the inbound flows into cells $j$ and $q$, respectively.  Note that all of the associated connector flows $y_{iu}^t, y_{pu}^t, y_{uv}^t, y_{vj}^t$, and $y_{vq}^t$ are binary variables in this modified representation.    
In this setup, separate cells for each of the turning movements proposed in previous work \cite{lin2004emi,beard2006sos} is not needed.  Our setup can also accommodate many different types of phasing schemes when considering signal timing; any two movements may proceed through the intersection simultaneously as long as they do not conflict. This is in contrast to the two phase setups in \cite{lo2001cbt}, \cite{lin2004emi} and \cite{beard2006sos}. Also, the binary variables $\alpha_{ij}^{t}$ representing the conflicting signal phases are no longer required to enforce conflict free flows at the intersection and we, henceforth, shall treat them as exogenous variables that are inferred from the other flow variables.  We also note that approaches with different capacities (approaches with different numbers of lanes) can also be accommodated in this framework as illustrated in the simple example shown in Fig.~\ref{fig:2lane}.

\subsection{Super graph: A static representation of traffic dynamics} \label{subsec:supergraph} 
We now demonstrate how the equations describing the system dynamics and the conflicting flows can be captured with a time-expanded static graph, which we refer to as the \emph{super-graph}.

\begin{figure}[h!]
	\centering
	\subfloat[][Cellular representation]{\resizebox{0.44\textwidth}{!}{
			\includegraphics[width=0.44\textwidth,origin=c]{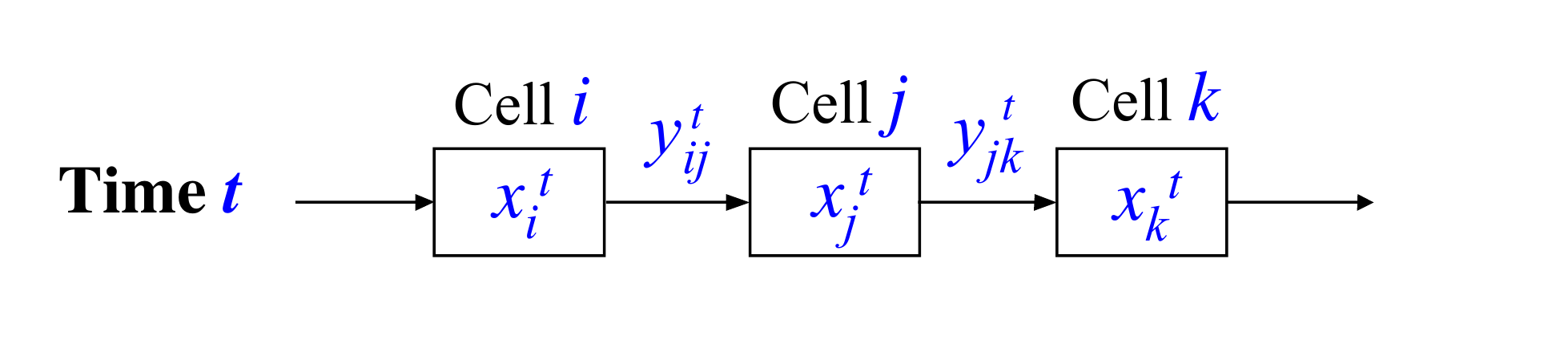}} 
		\label{fig:sg-illustra}} 	
	
	\subfloat[][Graphical representation of mass-balance equations]{\resizebox{0.44\textwidth}{!}{
			\includegraphics[width=0.44\textwidth,origin=c]{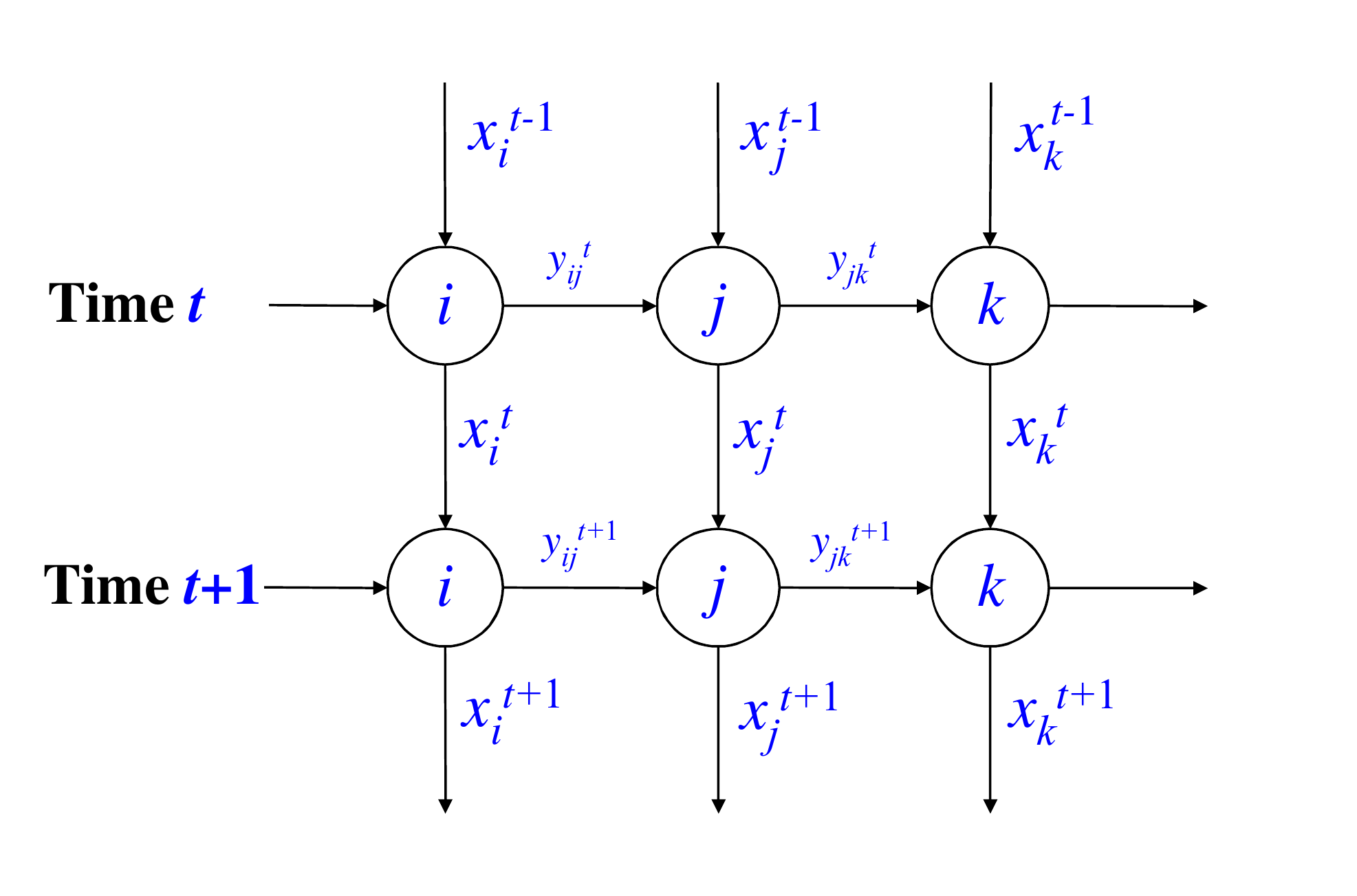}} 
		\label{fig:sg-illustrb}}
	
	\subfloat[][Graphical representation including flow restriction]{\resizebox{0.44\textwidth}{!}{
			\includegraphics[width=0.44\textwidth,origin=c]{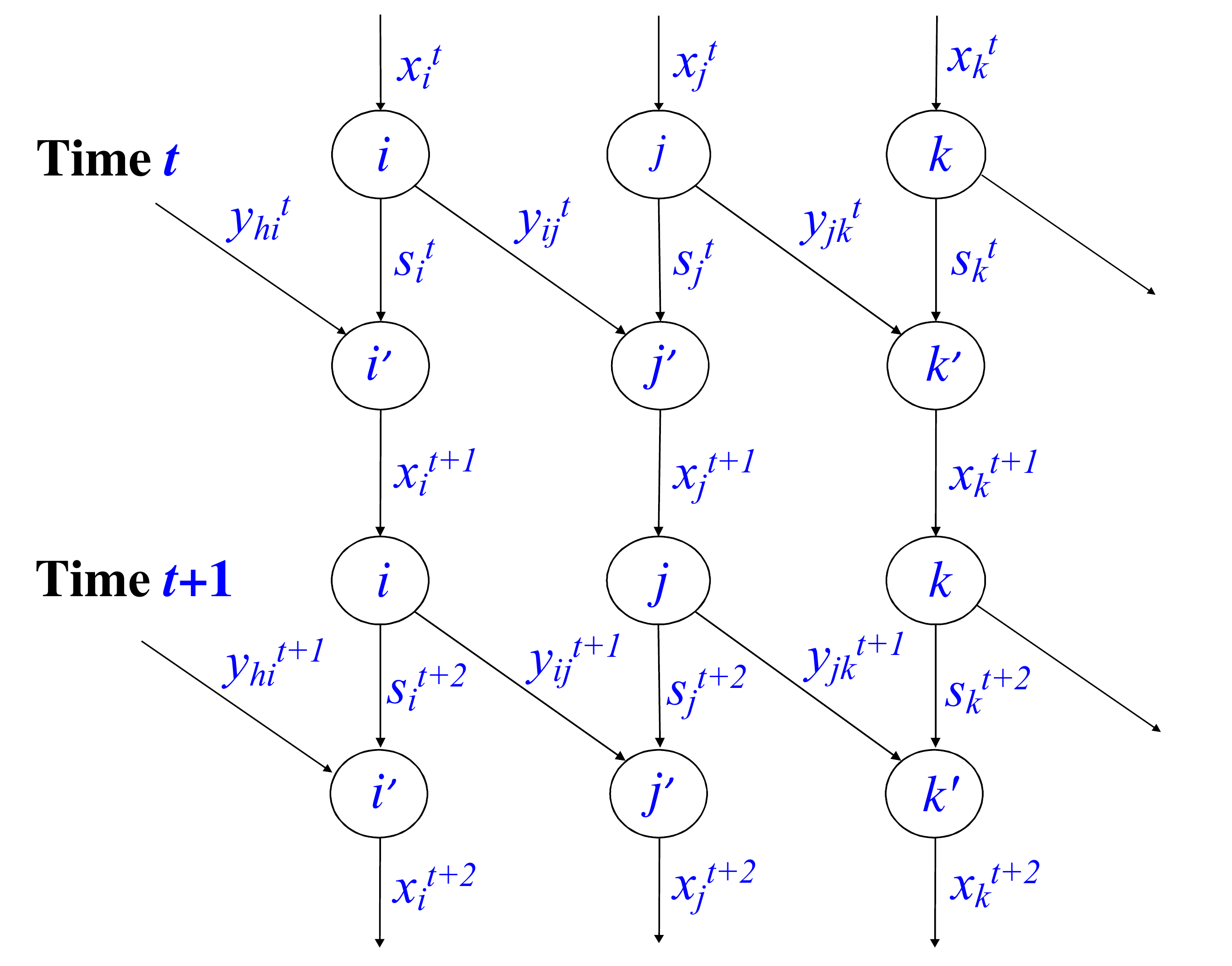}} 
		\label{fig:sg-illustrc}}
	\caption{\small Illustration of the time-expanded representation of the traffic dynamics.}
	\label{fig:sg-illustr}
\end{figure}

To illustrate, consider three adjacent cells $i, j, k$ in the network as depicted in Fig.~\ref{fig:sg-illustra}. The mass-balance equations governing the flow dynamics can be interpreted as flows through a network node. Take, for instance, the variable $x_j^t$ representing the occupancy of cell $j$ in time step $t$.  This variable appears twice in the system of mass-balance equations:
\begin{equation}
x_j^t - x_j^{t-1} - y_{ij}^{t-1} + y_{jk}^{t-1} = 0
\end{equation}
and in the next time step
\begin{equation}
x_j^{t+1} - x_j^t - y_{ij}^t + y_{jk}^t = 0. \label{massEx1}
\end{equation}
The variable $x_j^t$ appears in the first equation with a coefficient of +1 and in the second equation with a coefficient of -1.  It can hence be interpreted as \emph{temporal flow} from a node in time step $t$ into another copy in time step $t+1$.  Similarly take the variable $y_{ij}^t$, in addition to its appearance in equation \eqref{massEx1} with a coefficient of -1, it also appears in the mass balance equation
\begin{equation}
x_i^{t+1} - x_i^t - y_{hi}^t + y_{ij}^t = 0
\end{equation}
with a coefficient of +1.  It can be interpreted as a \emph{spatial flow}.  It can be easily seen that all cell occupancy variables $\{x_i^t\}_{i \in \CX, t \in \TX}$ and all connector flow variables $\{y_{ij}^t\}_{(i,j) \in \LX, t \in \TX}$ appear twice in the set of mass-balance equations, once with a coefficient of +1 and once with a coefficient of -1.  The time expanded directed graph associated with the mass balance equations of cells $i,j$ and $k$ is depicted in Fig.~\ref{fig:sg-illustrb}.

To incorporate the flow conservation constraints in the super-graph, we first introduce non-negative slack variables, denoted by $s_i^t$ for each cell $i$, which represents the number of vehicles in cell $i$ that do not advance to neighboring cell $j$ during $t$. Hence, they are bounded from above by the maximum occupancy of cell $i$:
\begin{equation} 
s_i^t  \le  N_i \label{eqn:slack}
\end{equation}
and the flow restrictions (\ref{eqn_flow_res_b}) becomes an equality:
\begin{equation}
y_{ij}^t - x_i^t + s_i^t = 0. \label{eqn:consa}
\end{equation}
Substituting $y_{ij}^t - x_i^t = -s_i^t$ in the mass balance equation
\begin{equation}
x_i^{t+1} - x_i^t - y_{hi}^t + y_{ij}^t = 0, \label{eqn:consb}
\end{equation} 
we obtain
\begin{equation}
x_i^{t+1} - y_{hi}^t - s_i^t = 0. \label{eqn:consaltb}
\end{equation} 
Then keeping \eqref{eqn:consa} in tact and replacing \eqref{eqn:consb} with \eqref{eqn:consaltb}, we have that each of the variables in the resulting system of equations $\{x_i^t,s_i^t\}_{i \in \CX, t \in \TX}$ and $\{y_{ij}^t\}_{i \in \LX, t \in \TX}$ appears twice, once with a coefficient of +1 and once with a coefficient of -1.  The resulting graph for the three-cell example is illustrated in Fig.~\ref{fig:sg-illustrc}.  
\begin{figure}[h!]
	\centering
	\subfloat[][Cell-node representation]{\resizebox{0.385\textwidth}{!}{
			\includegraphics[width=0.385\textwidth,origin=c]{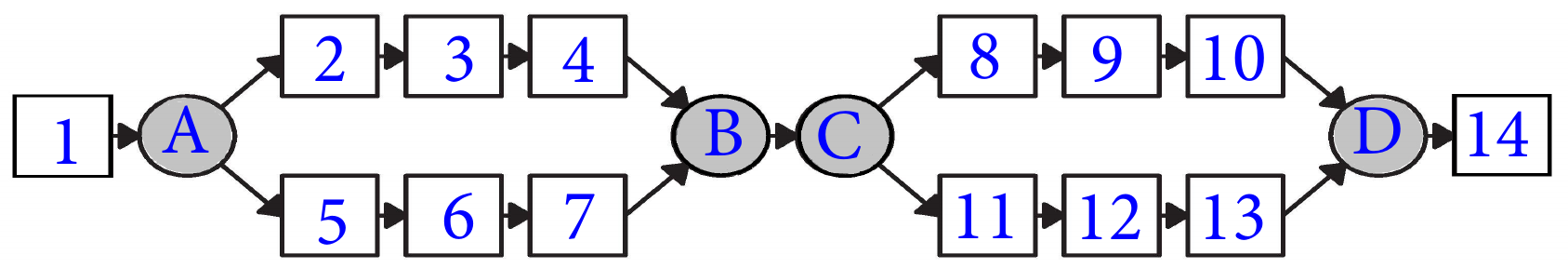}} 
		\label{fig:sg-exampa}} 	
	
	\subfloat[][A single layer of the super-graph]{\resizebox{0.49\textwidth}{!}{
			\includegraphics[width=0.49\textwidth,origin=c]{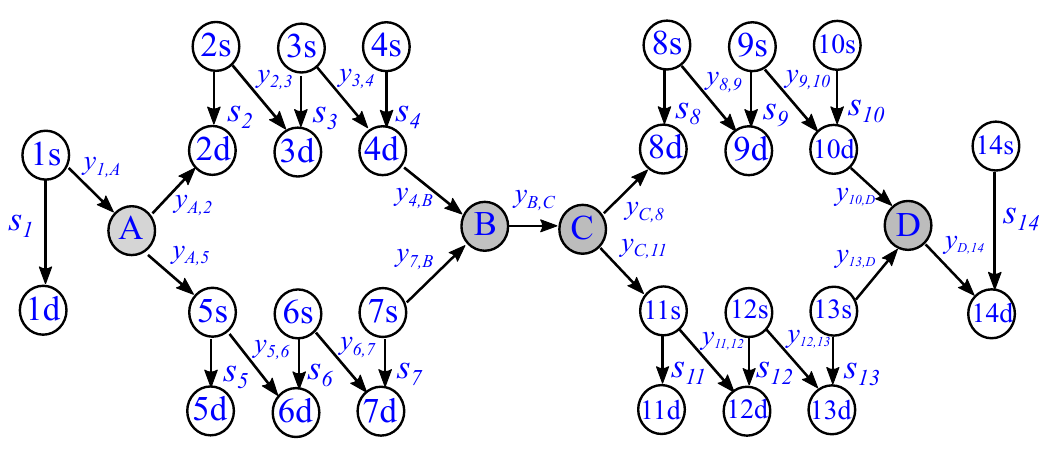}} 
		\label{fig:sg-exampb}}
	
	\subfloat[][Super-graph]{\resizebox{0.49\textwidth}{!}{
			\includegraphics[width=0.49\textwidth,origin=c]{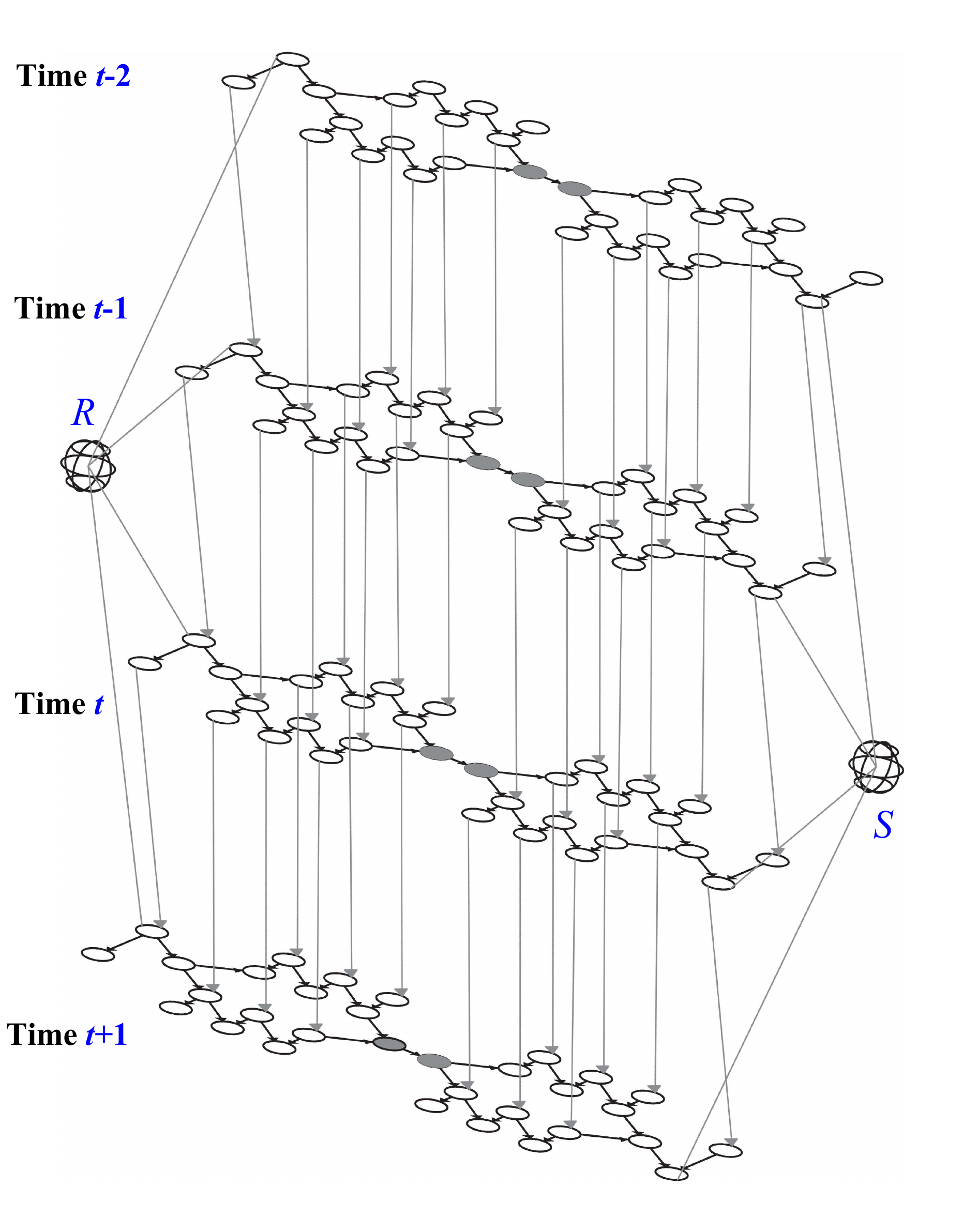}} 
		\label{fig:sg-exampc}}
	\caption{\small Example super-graph generation for a network with cells and transshipment nodes.  The super-graph consists of four layers representing four times steps}
	\label{fig:sg-examp}
\end{figure}

To fully capture the traffic dynamics, it remains to  accommodate downstream restriction constraints on the flows \eqref{eqn_flow_res_d}.  We do this by imposing a simple spatial capacity (max occupancy) constraint, which is natural:
\begin{equation}
x_i^t \le N_i \label{eqn_jam}
\end{equation}
for all $i \in \CX$ and all $t \in \TX$.  That the resulting system of equations respects \eqref{eqn_flow_res_d} by simply imposing \eqref{eqn_jam} follows by noting that for any $t$, mass balance implies that $x_j^{t+1} = x_j^t + y_{ij}^t - y_{jk}^t$ and since the right-hand side of \eqref{eqn_jam} is constant throughout the horizon $\TX$, we have that $x_j^{t+1} \le N_j$ implies that $x_j^t + y_{ij}^t - y_{jk}^t \le N_j$, which is the flow restriction constraint  \eqref{eqn_flow_res_d}.

We illustrate the overall development of the super-graph for a network with both cells (road segments) and transshipment nodes (intersections) in Fig.~\ref{fig:sg-exampa}.  A single layer of the supergraph corresponding to a single time step is shown in  Fig.~\ref{fig:sg-exampb}. The demands $\{d_i^t\}_{i \in \CR,t\in\TX}$ can be accommodated by adding a fictitious \emph{super-source} node $R$ along with fictitious arcs emanating from $R$ and terminating in the different copies of the source nodes with capacities equal to the demands.  The same is done for the network sinks, where we create a \emph{super-sink} node $S$ and connect it to the sink nodes via arcs that are either uncapacitated (for free-boundaries) or are capacitated by the maximum occupancies of the sink cells of the network.  The super-graph is illustrated in Fig.~\ref{fig:sg-exampc}.

In summary the network dynamics can be represented by a system of mass balance equations along with capacity constraints \eqref{eqn_flow_res_c}, \eqref{eqn:slack}, and \eqref{eqn_jam}.  

\section{Optimal and Adversarial Control}
\label{sec:failure-model}

\subsection{Compact Notation}
In this section, we formulate the optimal and adversarial control problem and show how the super-graph representation of traffic flow dynamics can be exploited in solving these problem efficiently.  We first present a compact notation based on the super-graph developed above.  
We denote the super-graph by $\GX = (\VX,\AX)$, where $\VX$ are the super-graph nodes (vertices) and $\AX$ are the directed arc in the super-graph. We denote by $\AXI \subset \AX$ the set of non-sink arcs in the network (in essence $\AXI$ excludes the fictitious arcs that terminate in the super-sink node $S$).  The subset of arcs corresponding to conflicting movements is denoted by $\AXJ \subset \AX$.  Let $\mathbf{b} \in \ZZ^{|\VX|}$ denote the demand vector.  Define $D \equiv \sum_{t \in \TX,i\in\CR} d_i^t$, then the element of the demand vector are given by
\begin{equation}
\mathbf{d}_i = \begin{cases}
D & i = R \\
-D & i = S \\
0 & \mbox{otherwise}
\end{cases}. \label{eqn_demands}
\end{equation}
The arc flows consist of the variables $\{x_i^t,s_i^t\}_{i \in \CX, t \in \TX}$ and $\{y_{ij}^t\}_{i \in \LX, t \in \TX}$, which for simplicity we denote as the vector $\mathbf{x} \in \ZZ^{|\AX|}$.  Let $\mathbf{u} \in \ZZ^{|\AX|}$ denote the upper-bounds on the flows, which correspond to the spatial capacities $\{N_i\}_{i \in \CX}$ and the temporal flow capacities $\{Q_i\}_{i \in \LX}$.  The mass-balance equations are linear equations in the elements of the vector $\mathbf{x}$ and can hence be represented by a matrix operator on $\mathbf{x}$, which we denote by $\mathbf{A} \in \{-1,0,1\}^{|\VX| \times |\AX|}$.  The mass balance equations are then written as $\mathbf{A}\mathbf{x} = \mathbf{b}$ along with the bounds $\mathbf{0} \le \mathbf{x} \le \mathbf{u}$.  The matrix $\mathbf{A}$ is an arc-node \emph{incidence matrix}.  Such matrices are known to possess a property called \emph{total unimodularity} (TU) \cite{wols1998integerProg,ahuja1993netflows}.  In the context of mixed integer programming, when the constraints matrix is TU, the trivial linear relaxation produces integer optimal solutions.  This is a property that we will capitalize on in the next section.

\subsection{Optimal control policy} 
\label{subsec:opt-contrl}
We define the optimal signal control policy as one that minimizes the total travel time of all vehicles in the system over the horizon of the problem.  This is modeled as $\Delta \tau \mathbf{1}_{|\AXI|}^{\top} \mathbf{x}|_{\AXI}$, where $\mathbf{1}_{|\AXI|}$ is a vector of ones of size $|\AXI|$ and $\mathbf{x}|_{\AXI}$ is the restriction of the vector $\mathbf{x}$ to arcs that lie in $\AXI$.  Since all conflicts are accounted for in $\GX$ naturally,the optimal control policy can be inferred from the solution of the network flow problem
\begin{equation}
\underset{\mathbf{x} \in \ZZ_+^{|\AX|}}{\mathrm{minimize}} ~ \big\{ \mathbf{1}_{|\AXI|}^{\top} \mathbf{x}|_{\AXI}: ~ \mathbf{A}\mathbf{x} = \mathbf{b}, ~\mathbf{x} \le \mathbf{u}\big\}, \label{eqn_SO}
\end{equation}
where one picks the solution of \eqref{eqn_SO} that involves the minimal number of signal switches.  Note that the constant $\Delta \tau$ was dropped from the objective function as it has no effect on the solution (since it is a positive constant). In the next section, we will denote by $\mathbf{x}^*$ the solution of \eqref{eqn_SO} and $\mathbf{x}_{\mathsf{conf}}^* \in \{0,1\}^{|\AXJ|}$ will denote the restriction of $\mathbf{x}^*$ to the set $\AXJ$.

\subsection{Adversarial signal control policy} 
\label{subsec:adv-contrl}

The signal tampering problem is formulated as a bi-objective optimization program, where we simultaneously seek to determine a control policy that leads to maximum adversarial impact in the network but at minimum noticeability. We call the resulting policy the \emph{adversarial} signal control policy and denote by $\mathbf{x}_{\mathsf{conf}}^{\mathsf{A}}\in \{0,1\}^{|\AXJ|}$ the intersection flows that result from the adversarial policy. Let $\mathbf{x}_S^*$ and $\mathbf{x}_S^{\mathsf{A}}$ denote the vector of flows into the sink node under optimal control and adversarial control, respectively.  The adversarial control policy can be inferred from the solution of the bi-objective programming problem
\begin{equation}
\mathbf{x}_S^{\mathsf{A}}  \in \underset{\mathbf{x} \in \ZZ_+^{|\AX|}}{\arg \min} ~ \big\{ Z_1(\mathbf{x}_S), ~Z_2(\mathbf{x}_{\mathsf{conf}}): ~ \mathbf{A}\mathbf{x} = \mathbf{b}, ~\mathbf{x} \le \mathbf{u}\big\}, \label{eqn_Adv}
\end{equation}
where the objective functions $Z_1: \AXI \rightarrow \ZZ_+$ and $Z_2: \AXJ \rightarrow \ZZ_+$ are given by
\begin{equation}
Z_1(\mathbf{x}_S) = - \| \mathbf{x}_S^* - \mathbf{x}_S \|_1
\end{equation}
and
\begin{equation}
Z_2(\mathbf{x}_{\mathsf{conf}}) = \| \mathbf{x}_{\mathsf{conf}}^*  - \mathbf{x}_{\mathsf{conf}} \|_0.
\end{equation}
The function $\|\cdot\|_1$ is the usual $\ell_1$-norm and $\| \cdot \|_0$ is the 0 pseudo-norm.  We assume that $\mathbf{x}_S^*$ and $\mathbf{x}_{\mathsf{conf}}^*$ are known \emph{a-priori}.  Hence, minimizing $Z_1$ can be interpreted as maximizing the deviation in network travel times from the one achieved under optimal control.  Minimizing $Z_2$ can be interpreted as finding a control strategy which results in smallest change in control variables from that achieved with optimal control.  Hence, the bi-objective programming problem can interpreted as one that seeks to find a control policy that impacts performance greatly while simultaneously going unnoticed.

The first objective can be simplified to 
\begin{equation}
Z_1(\mathbf{x}_S) = - (\mathbf{x}_S^* - \mathbf{x}_S  ) = \mathbf{x}_S  - \mathbf{x}_S^*.
\end{equation}
This follows from the equivalence of the system travel time minimizing solution and the maximal flow solution (see, e.g., \cite{ahuja1993netflows}).  Hence the cumulative vehicle arrivals (over time) curve under a optimal control solution lies above  any other cumulative arrivals curve.

The second objective, $Z_2$, is generally non-convex but noting that (feasible) flows in $\AXJ$ are binary under the proposed representation for conflicting movements, we have the convex equivalent:
\begin{equation}
Z_2(\mathbf{x}_{\mathsf{conf}}) = \| \mathbf{x}_{\mathsf{conf}}^*  - \mathbf{x}_{\mathsf{conf}} \|_0 = \| \mathbf{x}_{\mathsf{conf}}^*  - \mathbf{x}_{\mathsf{conf}} \|_1.
\end{equation}
To further simplify this, we employ a standard trick used in linear programming: introduce two binary vectors denoted $\mathbf{x}_{\mathsf{conf}}^+ \in \{0,1\}^{\AXJ}$ and $\mathbf{x}_{\mathsf{conf}}^-\in \{0,1\}^{\AXJ}$ and add the system of equations
\begin{equation}
\mathbf{x}_{\mathsf{conf}} = \mathbf{x}_{\mathsf{conf}}^+ - \mathbf{x}_{\mathsf{conf}}^- \label{eqn_trick}
\end{equation}  
to the system.  The addition of these constraints are equivalent, from the graph representation perspective, to every flow in the network corresponding to an element of $\mathbf{x}_{\mathsf{conf}}$ being replaced by an identical element of $\mathbf{x}_{\mathsf{conf}}^+$ and a new \emph{mirror image} in $\mathbf{x}_{\mathsf{conf}}^+$.  By mirror image, what is meant here is a flow that connects the same two nodes with the same capacity but points in the opposite direction.  Hence, this substitution has no impact on the graph structure of the problem or the total unimodularity of the constraints.

As far as the objective function $Z_2$ is concerned, the equality given in \eqref{eqn_trick} results in the following bound:
\begin{multline}
\| \mathbf{x}_{\mathsf{conf}}^*  - \mathbf{x}_{\mathsf{conf}} \|_1 = \| \mathbf{x}_{\mathsf{conf}}^*  - \mathbf{x}_{\mathsf{conf}}^+ + \mathbf{x}_{\mathsf{conf}}^- \|_1 \\
\le \mathbf{x}_{\mathsf{conf}}^*  + \mathbf{x}_{\mathsf{conf}}^+ + \mathbf{x}_{\mathsf{conf}}^-,
\end{multline}
which means that, without loss of generality, $Z_2$ can be replaced by the linear objective function:
\begin{equation}
Z_2(\mathbf{x}_{\mathsf{conf}}^+,\mathbf{x}_{\mathsf{conf}}^-) = \mathbf{x}_{\mathsf{conf}}^*  + \mathbf{x}_{\mathsf{conf}}^+ + \mathbf{x}_{\mathsf{conf}}^-.
\end{equation}
The resulting bi-objective programming problem is 
\begin{align}
\underset{\mathbf{x}, \mathbf{x}_{\mathsf{conf}}^+,\mathbf{x}_{\mathsf{conf}}^- }{\mathrm{minimize}} & \quad Z_1(\mathbf{x}_S) \label{eqn_advObj1} \\
\underset{\mathbf{x}, \mathbf{x}_{\mathsf{conf}}^+,\mathbf{x}_{\mathsf{conf}}^- }{\mathrm{minimize}} & \quad Z_2(\mathbf{x}_{\mathsf{conf}}^+,\mathbf{x}_{\mathsf{conf}}^-) \\
\mathrm{s.t.} & \quad \mathbf{A}\mathbf{x} = \mathbf{b}, \label{eqn_advEQ} \\
& \quad \mathbf{x}_{\mathsf{conf}} - \mathbf{x}_{\mathsf{conf}}^+ + \mathbf{x}_{\mathsf{conf}}^- = \mathbf{0}, \\
& \quad \mathbf{x} \le \mathbf{u}, \\
& \quad \mathbf{x} \in \RR_+^{|\AX|}, ~\mathbf{x}_{\mathsf{conf}}^+,\mathbf{x}_{\mathsf{conf}}^+ \in [0,1]^{|\AXJ|}. \label{eqn_advVT}
\end{align}
Note that we employed the trivial linear relaxation in \eqref{eqn_advVT}.  This is warranted by the total unimodularity of the constraint set, that is, we are still guaranteed integer optimal solutions.  This is true for each of the objective functions coupled with the constraints.  Thus, it is also true for any weighted combination of the two objective functions, which is the standard approach to solving bi-objective programming problems.  There also exist traditional techniques that ensure that all efficient solutions are discovered (i.e., the entire Pareto-optimal frontier is determined) \cite{aneja1979bicriteria,tayi1986bicriteria}.  Such techniques involve iteratively solving single objective problem but adding the other objective to the constraint set.  In our case, this too does not violate total unimodularity.  For example, minimizing $Z_1$ subject to \eqref{eqn_advEQ} - \eqref{eqn_advVT} and $Z_2 = z_2$ for some given constant $z_2$, we have the original TU system with rows from an identity matrix added to them.  The latter operation is \emph{TU-preserving}.  The exact same argument is true in the case where one minimizes $Z_2$ subject to \eqref{eqn_advEQ} - \eqref{eqn_advVT} and $Z_1 = z_1$ for some constant $z_1$.  Hence, efficient (polynomial-time) techniques can be employed to iteratively solve the bi-objective programming problem  \eqref{eqn_advObj1} - \eqref{eqn_advVT} and obtain the entire Pareto-optimal frontier.  For the sake of completeness, we include the algorithm we employed below.  To simplify notation, we represent the constraints \eqref{eqn_advEQ} - \eqref{eqn_advVT} as the set $\mathcal{X}$ and with slight notation abuse, the vector $\mathbf{x}$ includes $x_{\mathsf{conf}}^+$ and $x_{\mathsf{conf}}^-$.
\begin{algorithm}[h!]
	\small
	\caption{Pareto-optimal frontier for adversarial problem}
	\label{alg:soln-algo}		
	\KwData{}
		
	\KwResult{The Pareto-optimal frontier $\mathcal{F}$}
		
	\textbf{Initialize}: 
	$ z_{1}^{[1]} \mapsfrom \min \{Z_1: \mathbf{x} \in \mathcal{X}\} $
	
	$ z_{2}^{[1]} \mapsfrom \min \{Z_2: Z_1 \mapsfrom z_{1}^{[1]}, \mathbf{x} \in \mathcal{X}\} $
	
	$ z_{2}^{[2]} \mapsfrom \min \{Z_2: \mathbf{x} \in \mathcal{X}\} $
	
	$ z_{1}^{[2]} \mapsfrom \min \{Z_1:  Z_2 \mapsfrom z_{2}^{[2]}, \mathbf{x} \in \mathcal{X}\}$
	
	\eIf{$(z_{1}^{[1]}, z_{2}^{[1]}) \ne (z_{1}^{[2]}, z_{2}^{[2]})$}{
	
	$\mathcal{K} \mapsfrom \{ [1, 2] \}$ \tcp{set of candidate vertices of Pareto frontier}
	
	$\mathcal{F} \mapsfrom \emptyset$ \tcp{set of final vertices of Pareto frontier}
	
	$k \mapsfrom 3$
	}{
	Stop
	}
	\While{$\mathcal{K} \ne \emptyset$}{
	
	Choose any $[r, s] \in \mathcal{K}$
	
	Calculate the weights:
	
	\hspace{0.2in} $w_1^{[r,s]} \mapsfrom | z_{2}^{[s]} - z_{2}^{[r]} |$
	
	\hspace{0.2in} $w_2^{[r,s]} \mapsfrom | z_{1}^{[s]} - z_{1}^{[r]} |$
	
	Solve the following two problems:
	
	\hspace{0.2in} $\bar{z} \mapsfrom \min\{ w_{1}^{[r,s]} Z_{1} + w_{2}^{[r,s]} Z_{2} : \mathbf{x} \in \mathcal{X} \}$
	
	\hspace{0.2in} $\overline{\mathbf{x}} \mapsfrom \arg\min \{ Z_1 : w_{1}^{[r,s]} Z_1 + w_2^{[r,s]} Z_2 = \bar{z}, \mathbf{x} \in \mathcal{X} \}$
	
	Calculate candidate solution:
	
	\hspace{0.2in} $(\bar{z}_1, \bar{z}_2) \mapsfrom \big(Z_1(\overline{\mathbf{x}}_S), Z_2(\overline{\mathbf{x}}_{\mathsf{conf}}^+,\overline{\mathbf{x}}_{\mathsf{conf}}^-) \big)$
	
	\eIf{$(\bar{z}_1, \bar{z}_2) = (z_1^{[r]}, z_2^{[r]})$ or $(\bar{z}_1, \bar{z}_2) = (z_1^{[s]}, z_2^{[s]})$}{
	
	$\mathcal{F} \mapsfrom \mathcal{F} \cup [r,s]$
	
	$\mathcal{K} \mapsfrom \mathcal{K} / [r,s]$
	}{
	
	$z_1^{[k]}, z_2^{[k]} \gets \bar{z}_1, \bar{z}_2$
	
	$\mathcal{K} \mapsfrom \mathcal{K} \cup \{ [r,k], [k,s] \}$
	
	$\mathcal{K} \mapsfrom \mathcal{K} / [r,s]$

	$k \gets k + 1$
	}
}
\end{algorithm}

\begin{figure}[hbt!]
	\centering
	\resizebox{0.48\textwidth}{!}{%
		\includegraphics{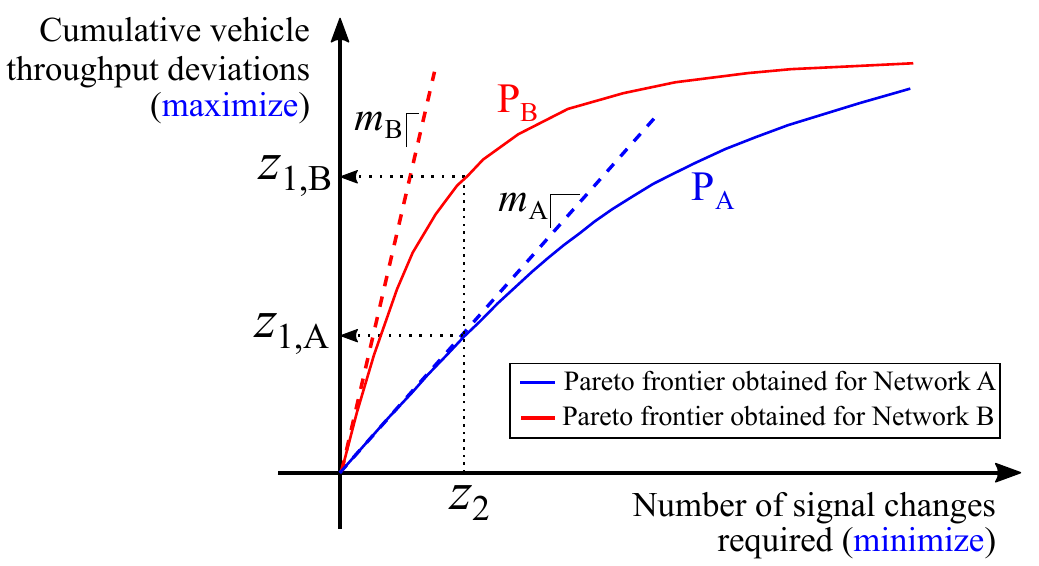} \hspace{0.2in}}
	\caption{ \small Pareto-optimal frontier for two hypothetical networks \textit{A} and \textit{B}.}
	\label{fig:net-vul}
\end{figure}

\textit{Pareto frontier and network vulnerability.} Each point along the Pareto-optimal frontier is a candidate solution which meets the dual criteria of maximum network vehicle throughput deviation and minimum noticeability.  Due to the discrete nature of the problem, the pareto frontier is scatter-plot (a set of extreme points that are not connected).  We shall illustrate the Pareto frontier as a piecewise linear curve (i.e., we will connect the points) to analyze the shape of the curve.  We examine the concavity of the resulting curve and use this as a measure of vulnerability of the network.  The more concave the curve is, the more vulnerable the network is to an adversarial attack.  
To visualize this, consider the two Pareto frontiers obtained for two different (hypothetical) networks $A$ and $B$ in Fig.~\ref{fig:net-vul}.
In Fig.~\ref{fig:net-vul}, an amount equivalent to $z_2$ change in the signal control results in a change in throughput equal to $z_{1,A}$ for network $A$ and $z_{1,B}$ for network $B$. Clearly, $z_{1,B} > z_{1,A}$, implying that network $B$ is more vulnerable than network $A$. Note here that the slope measured at the origin, which we will denote by $m_{\mathrm{A}}$ for network A, essentially tells how vulnerable the network is from an initial optimal state. The larger the value of $m$, the more vulnerable the network is to an attack.  One can also investigate slope and curvature at various points along the curve to determine the vulnerability and stability of the network to prolonged attacks.

\section{Numerical Experiments and Results}
\label{sec:numerical-expts}

\begin{figure}[h!]
	\centering
	\subfloat[][Network A]{\resizebox{0.245\textwidth}{!}{
			\includegraphics[width=0.245\textwidth,origin=c]{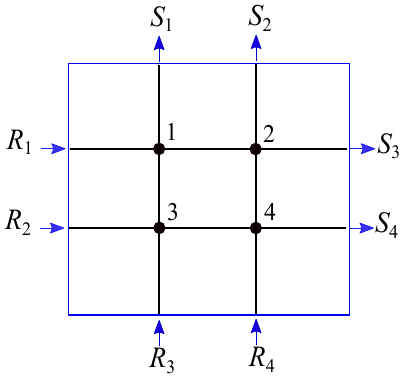}} 
		\label{fig:expt-netsa}} 	
	\subfloat[][Network B]{\resizebox{0.235\textwidth}{!}{
			\includegraphics[width=0.235\textwidth,origin=c]{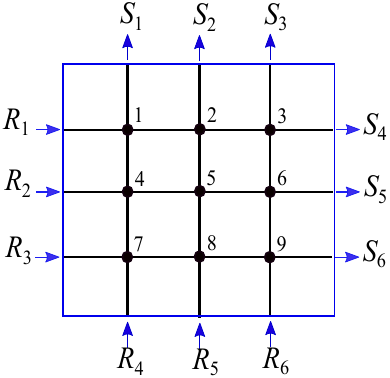}} 
		\label{fig:expt-netsb}}
	
	\subfloat[][Network C]{\resizebox{0.24\textwidth}{!}{
			\includegraphics[width=0.24\textwidth,origin=c]{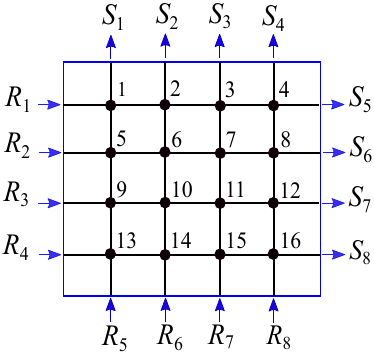}} 
		\label{fig:expt-netsc}}	
	\subfloat[][Network D]{\resizebox{0.24\textwidth}{!}{
			\includegraphics[width=0.24\textwidth,origin=c]{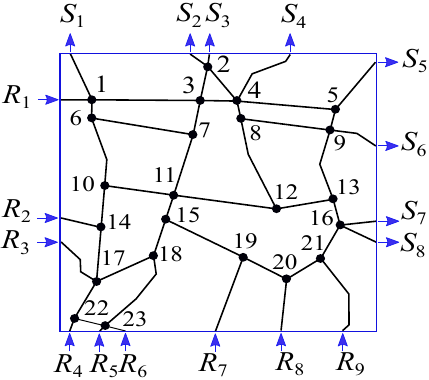}} 
		\label{fig:expt-netsd}}
	\caption{\small Test network layouts Network elements labeled `$R$' correspond to sources, network elements labeled `$S$' correspond to sinks.}
	\label{fig:expt-nets}
\end{figure}

We perform numerical experiments to test the proposed adversarial model on a few selected road networks, and discuss the dependence of network vulnerability on the network structure, vehicular traffic demand and duration of attack. We consider three homogeneous (or regular) road networks with uniform degree distributions and one random (or irregular) network with non-uniform degree distribution for the experiment; see Fig.~\ref{fig:expt-nets}. All of the networks in the examples have the same spatial area ($=2.5~\text{km}^2$), but vary in numbers of intersections, link lengths, origin (or destination) locations, and network regularity. These are detailed in Table \ref{tab:expt-net}. All the road links in the network have a single lane and only permit one-way traffic.
\begin{table*}[h!]
	\centering
	\caption{Characteristics of the test networks}
	\resizebox{0.7\textwidth}{!}{%
		\begin{tabular}{@{}lcccc@{}}
			\toprule
			\multicolumn{1}{c}{Network type} & \begin{tabular}[c]{@{}c@{}}Number \\ of intersections\end{tabular} & \begin{tabular}[c]{@{}c@{}}Number \\ of links\end{tabular} & \begin{tabular}[c]{@{}c@{}}Average \\ link  length\end{tabular} & \begin{tabular}[c]{@{}c@{}}Number of \\ source/sink points\end{tabular} \\ \midrule
			(A) Low density regular network & 4 & 12 & 500 & 4 \\
			(B) Medium density regular network & 9 & 24 & 375 & 6 \\
			(C) High density regular network & 16 & 40 & 250 & 8 \\
			(D) High density irregular network & 23 & - & 100 & 9/8 \\ \bottomrule
		\end{tabular}%
	}
	\label{tab:expt-net}
\end{table*}
For each network, we conduct a set of experiments with different traffic demands and attack durations.  
We assume a discrete time step length of $\Delta \tau = 2$ seconds, a maximum cell occupancy ($N_i$) of $5$ vehicles per cell, and a flow capacity of $Q_i = 1$ vehicle per time step per lane. These are typical values for an urban lane. The super-graph and bi-objective attack problem for the test networks are implemented using Python programming language and are available online \cite{github2019bilz}.  All linear optimization programs are solved using the \emph{Gurobi optimization solver} \cite{gurobi}. The experimental results are discussed below.

\begin{figure}[h!]
	\centering
	\subfloat[][Network A]{\resizebox{0.24\textwidth}{!}{
			\includegraphics[width=0.24\textwidth,origin=c]{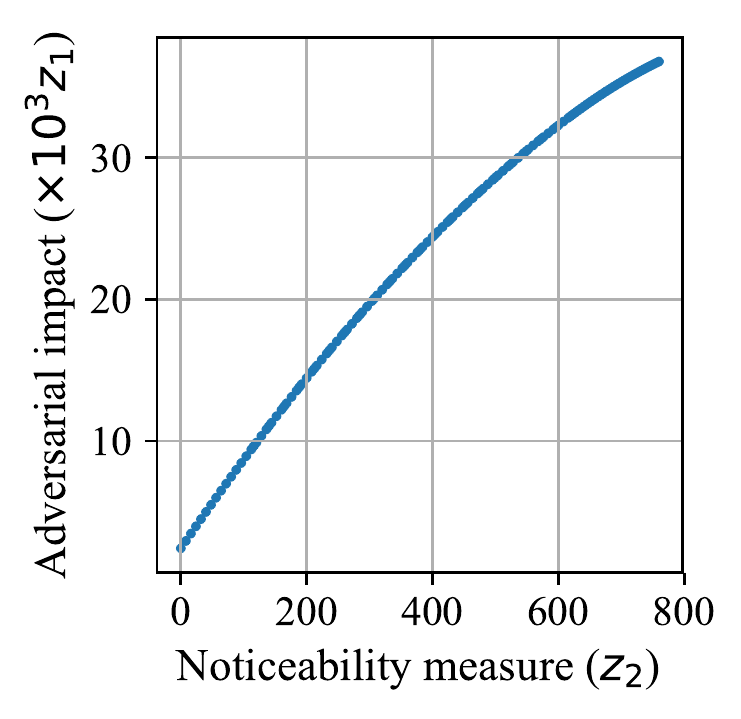}} 
		\label{fig:pareto-fronta}} 	
	\subfloat[][Network C]{\resizebox{0.24\textwidth}{!}{
			\includegraphics[width=0.24\textwidth,origin=c]{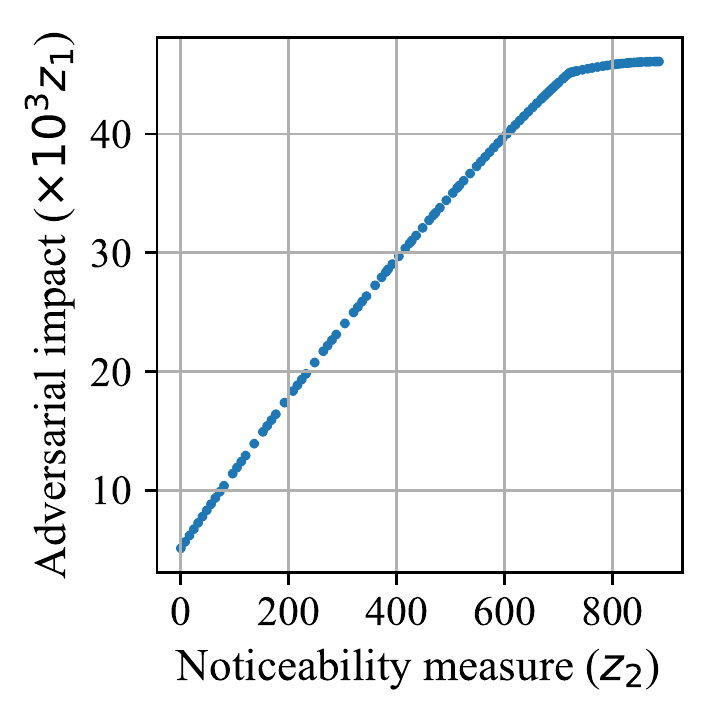}} 
		\label{fig:pareto-frontb}}
	\caption{\small Pareto-optimal frontiers for Network A and Network C.}
	\label{fig:pareto-front}
\end{figure}

\textit{Network vulnerability.} We first investigate network vulnerability in general by examining the concavity of the Pareto frontier in Fig.~\ref{fig:pareto-front} for the low-density and high-density networks under a uniform traffic demand of 600 vehicles per hour at the network sources (low traffic demands). 
The duration of the attack is 15 minutes (450 time steps).  We see a flatter curve in Fig.~\ref{fig:pareto-fronta} compared to Fig.~\ref{fig:pareto-frontb} indicating that the higher density network is more vulnerable in low demand scenarios.  This is reasonable as Network C has more intersections per unit area than Network A.



\begin{figure}[h!]
	\centering
	\subfloat[][Low traffic demands]{\resizebox{0.24\textwidth}{!}{
			\includegraphics[width=0.24\textwidth,origin=c]{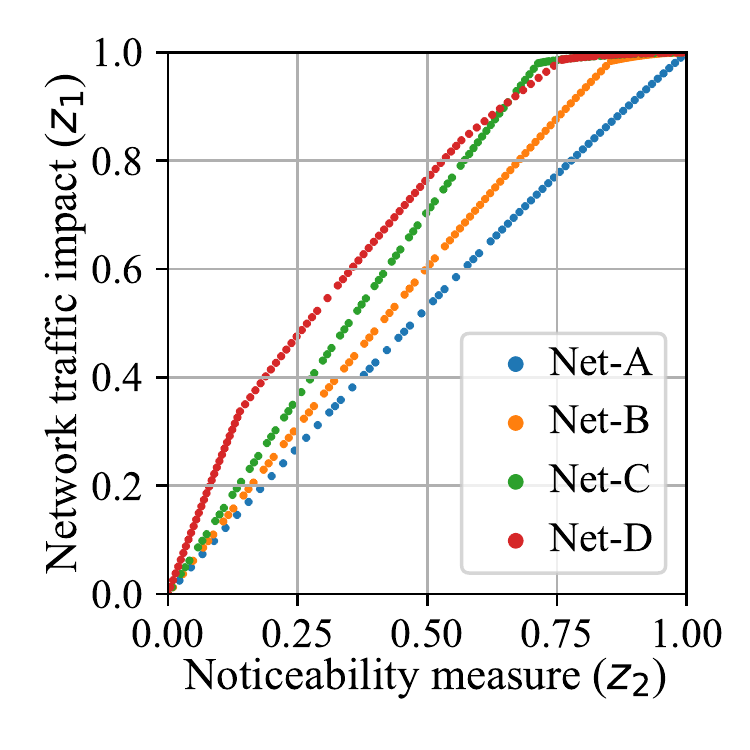}} 
		\label{fig:netstruct-compa}} 		
	\subfloat[][Moderate traffic demands]{\resizebox{0.24\textwidth}{!}{
			\includegraphics[width=0.24\textwidth,origin=c]{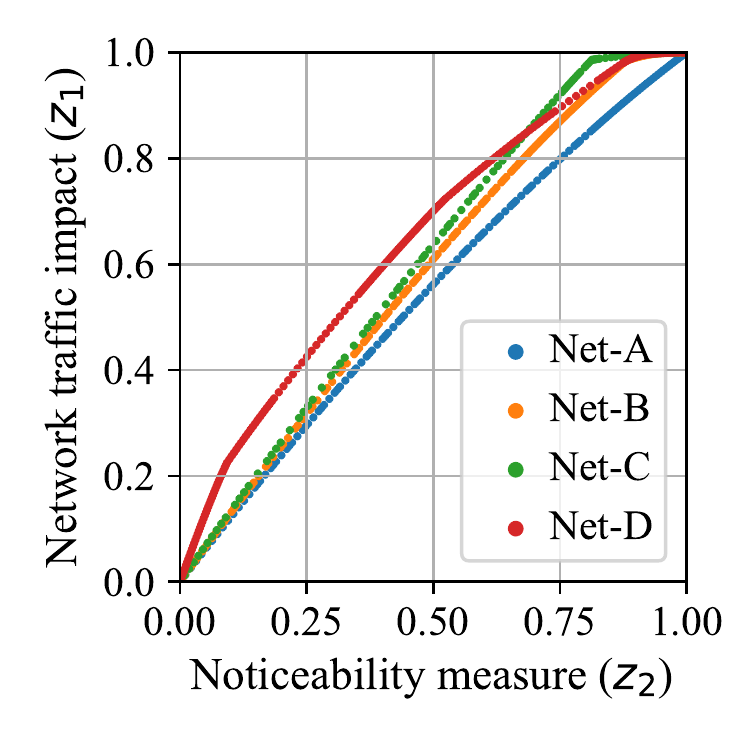}} 
		\label{fig:netstruct-compb}}
	
	\subfloat[][High traffic demans]{\resizebox{0.24\textwidth}{!}{
			\includegraphics[width=0.24\textwidth,origin=c]{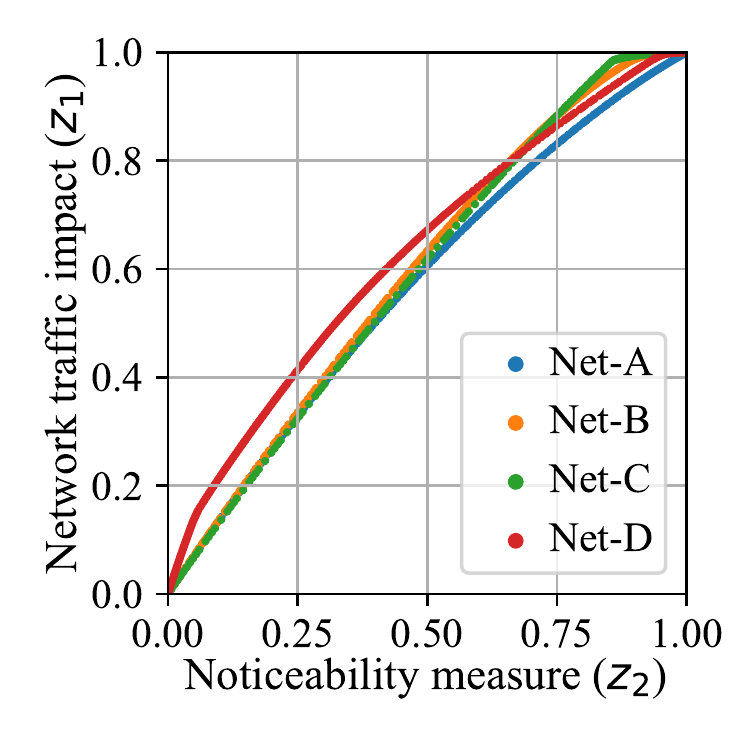}} 
		\label{fig:netstruct-compc}}
	
	\caption{\small Comparison of Pareto-optimal curves obtained for all test networks at various demand levels to see the relationship between network structure and network vulnerability. Sub-figures represent the comparison at different traffic demand levels (Blue: Network A, Orange: Network B, Green: Network C, Red: Network D).}
	\label{fig:netstruct-comp}
\end{figure}

\textit{Impact of network structure.} To understand the relationship between network structure and network vulnerability, we compare the Pareto frontiers (on a normalized scale) obtained for all the test networks in Fig.~\ref{fig:netstruct-comp}.  
The attack duration is 15 minutes (450 times steps) in the experiments. Fig.~\ref{fig:netstruct-compa} compares the networks under low traffic demands (400 veh/hr or 100 vehicles over the entire horizon at each source), Fig.~\ref{fig:netstruct-compb} compares the networks under moderate traffic demands (800 veh/hr or 200 vehicles over the entire horizon at each source), and Fig.~\ref{fig:netstruct-compc} compares the networks under heavy traffic demands (1200 veh/hr or 300 vehicles over the entire horizon at each source).  
In Fig.~\ref{fig:netstruct-compa}, we see that at the origin point, the slope values follow: $m_{\mathrm{D}} > m_{\mathrm{C}} > m_{\mathrm{B}} > m_{\mathrm{A}}$, implying the Network D is the most vulnerable and Network A is the least vulnerable in low demand scenarios. The high vulnerability of Network D could be attributed to its irregular structure and non-uniform degree distribution, for which there exist large degree nodes and are found to be more unstable under attacks. However, along the Pareto curve, the slopes reduce at a faster rate for Network D than Network A, implying that Network D tends to be more resilient than other networks under prolonged attacks. Similar insights can be observed from the Pareto curves obtained for the moderate and high demand scenarios in Fig.~\ref{fig:netstruct-compb} and Fig.~\ref{fig:netstruct-compc}.

\begin{figure}[h!]
	\centering
	\subfloat[][Network A]{\resizebox{0.24\textwidth}{!}{
			\includegraphics[width=0.24\textwidth,origin=c]{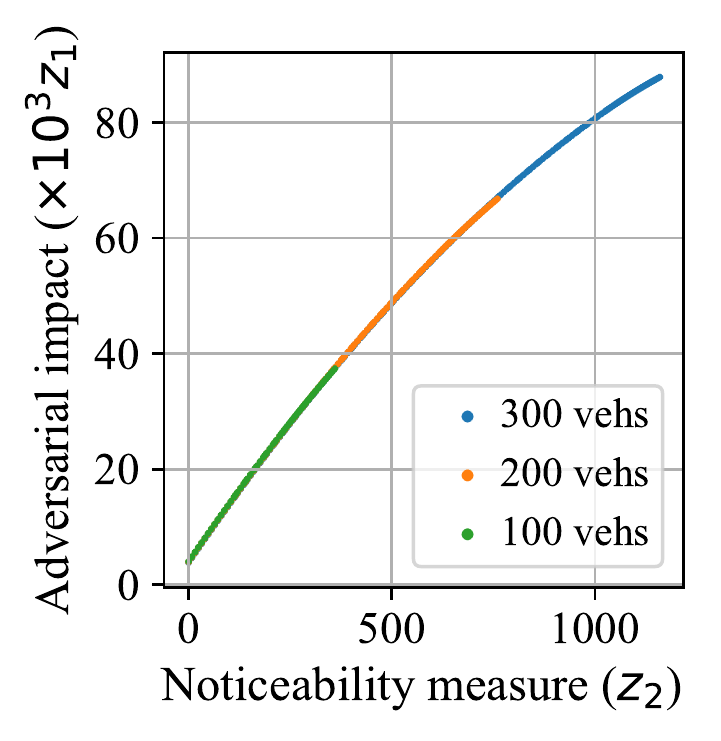}} 
		\label{fig:netdemnd-compa}} 		
	\subfloat[][Network B]{\resizebox{0.24\textwidth}{!}{
			\includegraphics[width=0.24\textwidth,origin=c]{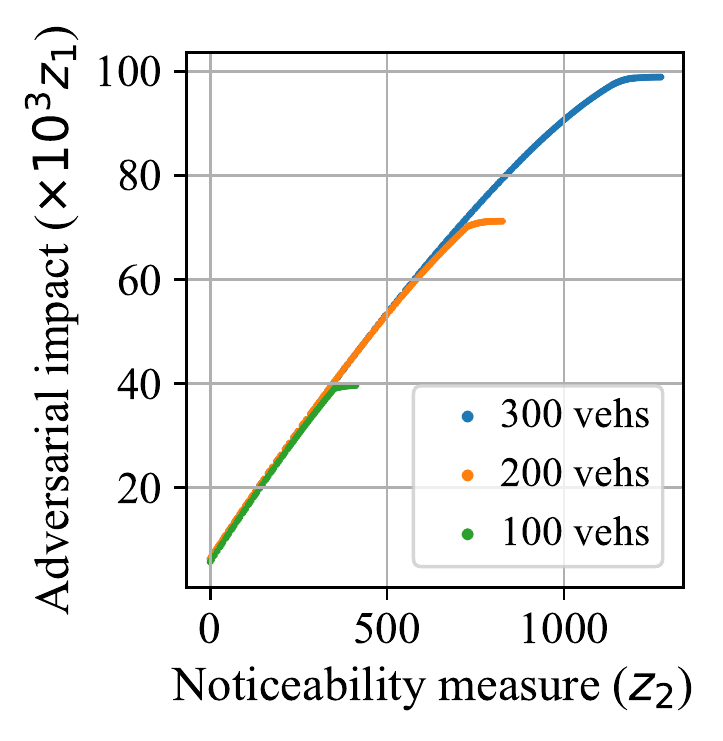}} 
		\label{fig:netdemnd-compb}}
	
	\subfloat[][Network C]{\resizebox{0.24\textwidth}{!}{
			\includegraphics[width=0.24\textwidth,origin=c]{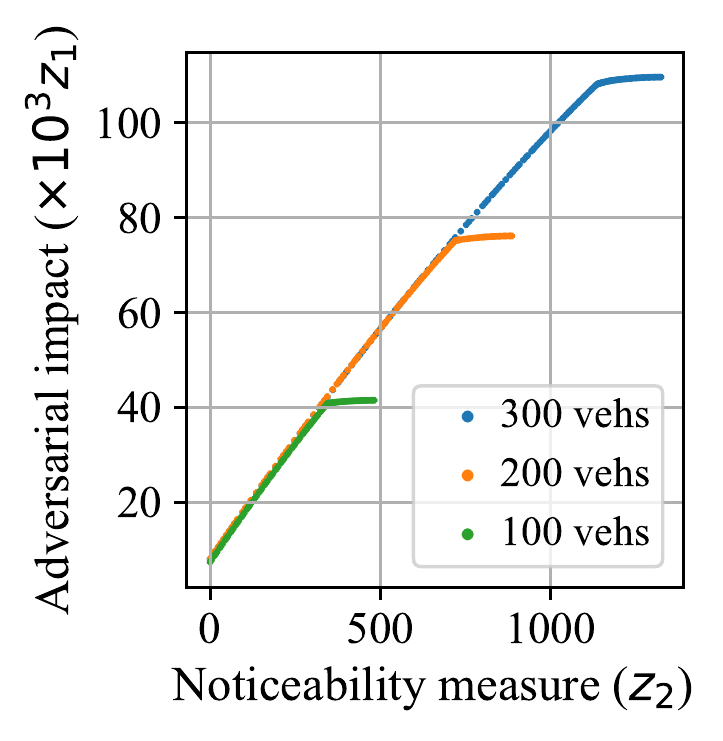}} 
		\label{fig:netdemnd-compc}}
	\subfloat[][Network D]{\resizebox{0.24\textwidth}{!}{
			\includegraphics[width=0.24\textwidth,origin=c]{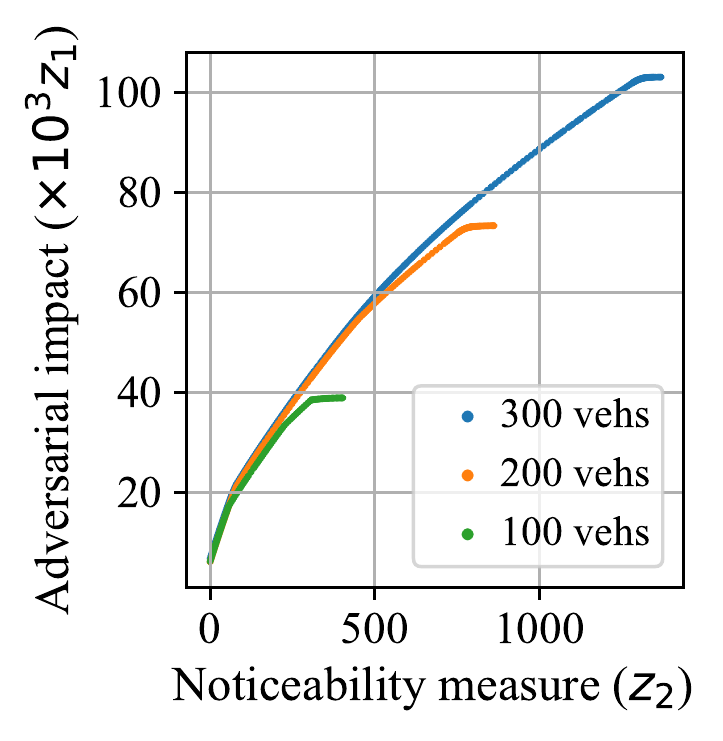}} 
		\label{fig:netdemnd-compd}}
	
	\caption{\small Illustration of network vulnerability with varying demand levels (Green: low demand scenario, Orange: moderate demand scenario, Blue: high demand scenario).}
	\label{fig:netdemnd-comp}
\end{figure}

\textit{Impact of  traffic demands.} We next study the effect of varying traffic demands; this is shown in Fig.~\ref{fig:netdemnd-comp}. 
For each network, we observe that the number of non-dominated solutions defining the Pareto frontier increases as demands increase.  The more interesting observation is the similar shapes of the curves across demand levels, particularly for the uniform networks (Networks A - C).  This similarity suggests that demand levels have no impact on a network's vulnerability, which is defined by the shape of the Pareto frontier.  We also see that the irregularity of Network D affords it less vulnerability to adversarial attacks than the uniform networks.

\begin{figure}[h!]
	\centering
	\subfloat[][Network A]{\resizebox{0.24\textwidth}{!}{
			\includegraphics[width=0.24\textwidth,origin=c]{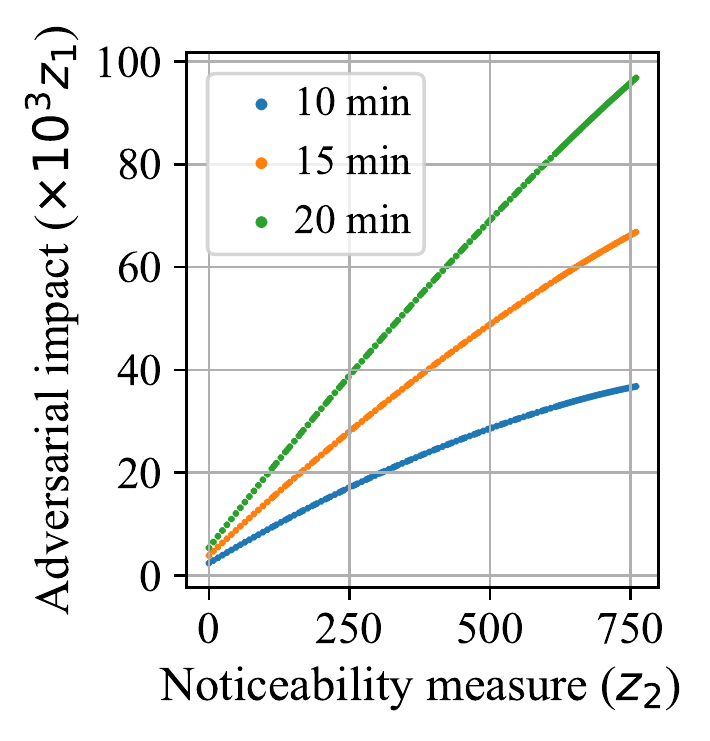}} 
		\label{fig:attacktime-compa}} 		
	\subfloat[][Network B]{\resizebox{0.24\textwidth}{!}{
			\includegraphics[width=0.24\textwidth,origin=c]{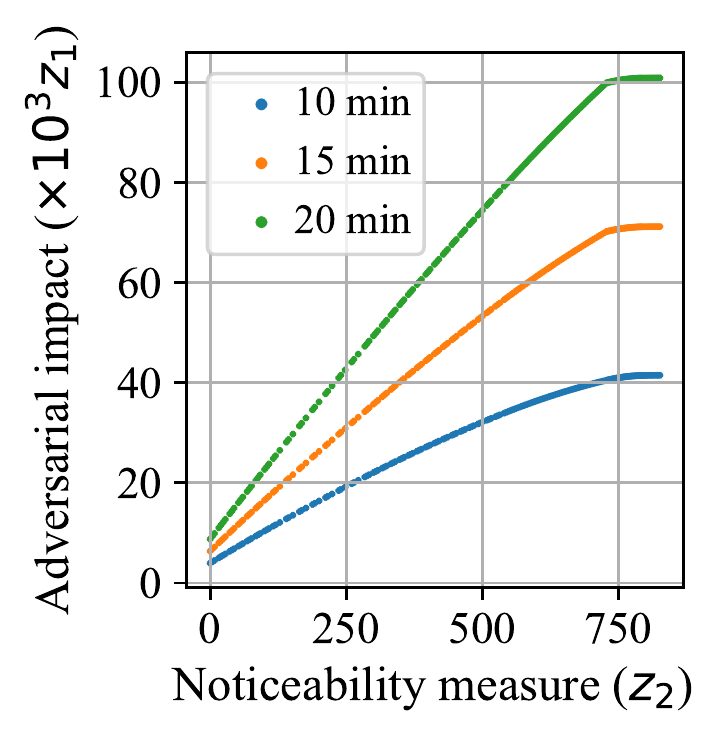}} 
		\label{fig:attacktime-compb}}
	
	\subfloat[][Network C]{\resizebox{0.24\textwidth}{!}{
			\includegraphics[width=0.24\textwidth,origin=c]{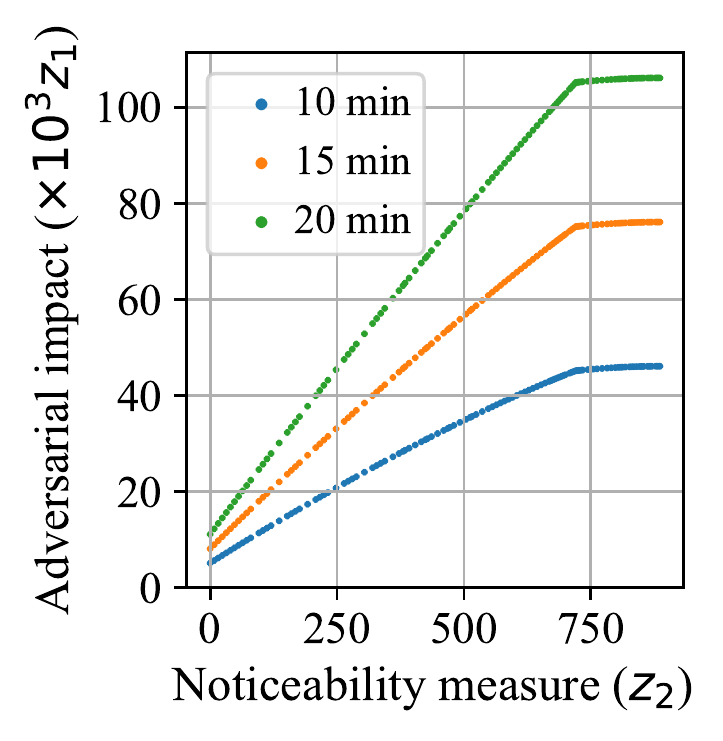}} 
		\label{fig:attacktime-compc}}
	\subfloat[][Network D]{\resizebox{0.24\textwidth}{!}{
			\includegraphics[width=0.24\textwidth,origin=c]{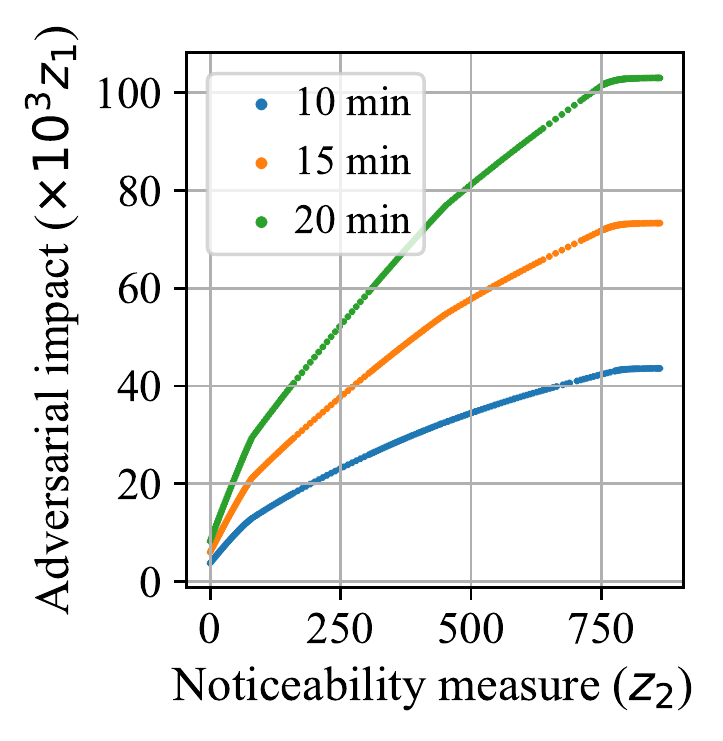}} 
		\label{fig:attacktime-compd}}
	
	\caption{\small Comparison of Pareto-optimal curves under different attack durations. (Green: 20 minutes = 600 time steps, Orange: 15 minutes = 450 timee steps, Blue: 10 minutes = 300 time steps).}
	\label{fig:attacktime-comp}
\end{figure}

\textit{Impact of attack duration.} Finally, Fig.~\ref{fig:attacktime-comp} illustrates the impact of attack duration. There are two noteworthy observations: the first observation is that all four networks are more vulnerable to longer attacks.  This is to be expected.  The second observation is that the Pareto curves scale exponentially with duration and it appears that this scaling is the same across the four networks.

\section{Conclusion}
\label{sec4:Concls}

A trend that is sweeping traffic signal systems worldwide is the move to wireless communication between sensors and controllers and between controllers and the signal heads. Communication between the intersections and the traffic management centers is also moving to wireless. This creates vulnerabilities to cyber attacks at various levels in the signal control system.  This paper considers the scenario in which an adversary gains access to the system and we model an attack that is intended to create congestion in the network with minimal changes to the signal timing plans.  The contributions of this paper are summarized as (i) the development of a representation of urban traffic dynamics as flows in a static directed (capacitated) graph, (ii) a model of the signal tampering attack as a bi-objective programming problem, and (iii) a demonstration that the problem can be solved to optimality using classical techniques. Furthermore, we provide examples of insights that can be extracted from the Pareto-optimal frontier that is obtained as a solution to our bi-objective problem.  Specifically, we examine the concavity of the curve as a measure of vulnerability of the urban traffic network.

Some of the observations that we make in our experiments suggest some scale-free properties of Pareto-optimal curves.  For example, we observe that the curves have the same shape under different demand levels.  We also observe an exponential scaling property with increasing attack durations that appears to be transferable from one network to another.  These observations suggest that the vulnerability of a network is an intrinsic property that is independent of demand level, attack duration, and in some cases (perhaps) network structure. These observations  motivate future research directions along these lines.

To the best of our knowledge, this is the first paper to treat the concavity of a Pareto-frontier as a way to gauge the vulnerability of a network.  This idea is transferable to any type of network in which there exist trade-offs between impact and noticeability, two broad notions that are always at odds with one another.

\section*{Acknowledgment}
This work was supported by the NYUAD Center for Interacting Urban Networks (CITIES), funded by Tamkeen under the NYUAD Research Institute Award CG001 and by the Swiss Re Institute under the Quantum Cities\textsuperscript{TM} initiative.

\appendix
\gdef\thesection{Appendix \Alph{section}}

	
	
	
\bibliographystyle{plainnat}
\bibliography{sigtamp-references.bib}
	
	
	
	
	
	

\end{document}